%% file: main4_2023.tex
\documentclass[a4paper,11pt]{article}

\usepackage{jheppub} 

\usepackage[T1]{fontenc}
\usepackage{multirow}
\usepackage{longtable}
\usepackage{graphicx}
\usepackage{dcolumn}
\usepackage{bm}
\usepackage{epsfig}
\usepackage{gensymb}
\usepackage{mathrsfs}
\usepackage{extpfeil}
\usepackage{extarrows}
\usepackage{float}
\usepackage{pdfpages}
\usepackage{amsmath}
\usepackage{amssymb}
\usepackage{graphicx,bm}
\usepackage{natbib}
\usepackage{slashed}
\usepackage[makeroom]{cancel}
\usepackage{dsfont}
\usepackage{longtable}
\usepackage{multirow}
\usepackage{youngtab}
\usepackage{young}
\usepackage{tikz}
\usepackage{cleveref}

\newcommand{\fd}{D}
\newcommand{\pd}{d}

\newcommand{\BH}{{\{\mathcal{B} \cdot H \}}}

\newcommand{\sbk}[1]{\left[{#1}\right]}
\newcommand{\abk}[1]{\left\langle{#1}\right\rangle}
 \usepackage[titletoc]{appendix}

\def\fun#1#2{\lower3.6pt\vbox{\baselineskip0pt\lineskip.9pt
  \ialign{$\mathsurround=0pt#1\hfil##\hfil$\crcr#2\crcr\sim\crcr}}}

\usepackage{tikz}
\usetikzlibrary{decorations.pathmorphing,decorations.markings,trees,shapes}

\def\lsim{\mathrel{\rlap{\raise 2.5pt \hbox{$<$}}\lower 2.5pt\hbox{$\sim$}}}
\def\gsim{\mathrel{\rlap{\raise 2.5pt \hbox{$>$}}\lower 2.5pt\hbox{$\sim$}}}

\input epsf

\newcommand{\blue}[1]{\textcolor{blue}{#1}}
\usepackage{color}

\newcommand{\onec}{1^{c}}
\newcommand{\twoc}{2^{c}}
\newcommand{\threec}{3^{c}}
\newcommand{\fourc}{4^{c}}
\newcommand{\fivec}{5^{c}}
\newcommand{\sixc}{6^{c}}
\newcommand{\sevenc}{7^{c}}
\newcommand{\eightc}{8^{c}}
\newcommand{\ninec}{9^{c}}

\newcommand{\comment}[1]{}

\newcommand{\be}{\begin{equation}}
\newcommand{\ee}{\end{equation}}
\newcommand{\bea}{\begin{eqnarray}}
\newcommand{\eea}{\end{eqnarray}}

\newcommand{\vev}[1]{\langle #1 \rangle}
\newcommand{\ket}[1]{| #1 \rangle}
\newcommand{\bra}[1]{\langle #1 |}
\newcommand{\antiket}[1]{| #1 ]}

\newcommand{\eq}[1]{\begin{equation}\begin{split} #1 \end{split}\end{equation}}

\renewcommand\arraystretch{2}
\makeatletter
\gdef\@fpheader{}
\makeatother

\begin{document}

\title{\boldmath The New Formulation of Higgs Effective Field Theory}

\author[a,b,c]{Zi-Yu Dong}
\author[d,e]{Teng Ma,}
\author[f,g]{Jing Shu}
\author[a,b]{Zi-Zheng Zhou}


\affiliation[a]{CAS Key Laboratory of Theoretical Physics, Institute of Theoretical Physics,
Chinese Academy of Sciences, Beijing 100190, China.}
\affiliation[b]{School of Physical Sciences, University of Chinese Academy of Sciences, Beijing 100190, P. R. China.}
\affiliation[c]{Department of Physics, LEPP, Cornell University, Ithaca, NY 14853, USA}

\affiliation[d]{IFAE and BIST, Universitat Aut\`onoma de Barcelona, 08193 Bellaterra, Barcelona}

\affiliation[e]{Physics Department, Technion -- Israel Institute of Technology, Haifa 3200003, Israel}

\affiliation[f]{School of Physics and State Key Laboratory of Nuclear Physics and Technology, Peking University, Beijing 100871, China}
\affiliation[g]{Center for High Energy Physics, Peking University, Beijing 100871, China}


\emailAdd{zd79@cornell.edu}
\emailAdd{mateng@ucas.ac.cn}
\emailAdd{jshu@pku.edu.cn}
\emailAdd{zhouzizheng@itp.ac.cn}

\abstract{
We present the explicit construction of the effective field theory (EFT) of standard model mass eigenstates. 
The EFT, which is invariant under $U(1)_{\text{e.m.}}\times SU(3)_c$, is constructed based on the on-shell method and Young Tableau technique. 
This EFT serves as a new formulation of the Higgs EFT (HEFT), which can describe the infrared effects of new physics at the electroweak symmetry-breaking phase with greater conciseness. The current HEFT operator basis has a clear physical interpretation, making it more accessible for research in phenomenology. 
A complete list of HEFT operator bases for any-point vertices up to any dimension could be provided, and three- and four-point bases are provided as examples.
Additionally, this framework realized as Mathematica program can be used to construct the EFT of any type of dark matter or particles with any spin.}


\maketitle
\section{Introduction} 

Effective field theories (EFTs) with massive fields find a broad range of applications in particle physics, including but not limited to lower energy quantum chromodynamics~\cite{Weinberg:1966kf, Weinberg:1968de, Weinberg:1978kz, Gasser:1982ap, Gasser:1983yg}, high spin particles~\cite{Hassan:2011zd, Rarita:1941mf, Fierz:1939ix}, and dark matter~\cite{Goodman:2010ku,Cao:2009uw,Zheng:2010js,Aebischer:2022wnl,DelNobile:2011uf,Babichev:2016bxi}. To fully explain the infrared (IR) effects of the generic ultraviolet (UV) theories, it is necessary to construct a complete set of EFT operator bases. However, this task is complicated because EFT operators relate to each other through equations of motion (EOMs) and integration by parts (IBP). A more efficient method of constructing massless EFT basis is to utilize the on-shell method~\cite{Shadmi:2018xan,Ma:2019gtx,Henning:2019enq, Falkowski:2019zdo,Durieux:2019siw}, which capitalizes on the one-to-one correspondence between an EFT operator and an on-shell amplitude basis. On-shell method is powerful to construct massless EFT basis because it is free of EOM redundancy~\cite{Shadmi:2018xan,Ma:2019gtx}. And the IBP redundancy can be systematically removed through the global $U(N)$ symmetry of massless spinors, where $N$ represents the number of external legs~\cite{Henning:2019enq}.
However, massive basis poses two additional challenges: EOM redundancy and possible overall mass factors that affect the base dimension.
The construction of 3-point and 4-point amplitudes for distinguishable particles is demonstrated in~ \cite{Durieux:2019eor, Durieux:2020gip}. In addition to that, we also take into account the effects of identical particles and gauge structures, presenting a systematic construction of the EFT operators for a general model.
A series of works~\cite{Dong:2021yak,Dong:2022mcv} have found that Lorentz subgroup $SU(2)_R$ and $U(N)$ can eliminate these redundancies and the lowest dimensional basis, which do not have the overall mass factors, can be obtained through the use of an algebraic method.
It is worth mentioning that \cite{DeAngelis:2022qco} proposed a different method for constructing massive amplitude basis.

Using the method described above, we have leveraged a Mathematica program to construct all $3$-pt and $4$-pt EFT operator bases of mass eigenstates of the standard model (SM) up to dimension-$8$ at the electroweak symmetry breaking (EWSB) phase. While our program can generate higher-point basis at any dimension, we have limited our focus to $3$-pt and $4$-pt processes in this study. Notably, this method is agnostic to how electroweak symmetry is broken, enabling the generation of a generic EFT of massive standard model fields that is invariant under $U(1)_{\text{EM}} \times SU(3)_c$. This massive HEFT can describe non-decoupling new physics beyond the SM, unlike traditional massless SMEFT.
In traditional HEFT~\cite{Feruglio:1992wf,Bagger:1993zf,Koulovassilopoulos:1993pw,Burgess:1999ha,Grinstein:2007iv,Alonso:2012px,Espriu:2013fia,Buchalla:2013rka,Brivio:2013pma,Alonso:2015fsp,Alonso:2016oah,Buchalla:2017jlu,Alonso:2017tdy,deBlas:2018tjm,Falkowski:2019tft,Cohen:2020xca,Sun:2022ssa,Sun:2022snw}, basis is constructed using the linearly realized sigma field of Goldstone and electroweak gauge eigenstates based on the Callan-Coleman-Wess-Zumino (CCWZ) theory~\cite{Coleman:1969sm,Callan:1969sn}. However, this basis contains unphysical Goldstone, impeding their physical interpretation and making the HEFT basis complex. Therefore, considering mass eigenstates permits HEFT to be formulated much simpler and more conveniently for phenomenological applications, while still accounting for the IR effects of UV theories at the EWSB phase concisely.
For instance, massive HEFT has fewer bases than traditional HEFT, since certain CCWZ bases that describe the same physical interaction only differ by the insertions of the Goldstone sigma fields. 
Besides, some CCWZ bases can indirectly affect some experimental measurements by correcting the wave functions or masses of some SM fields in the unitary gauge, making measurement parameters correlated and redundant. In contrast, massive HEFT does not have these issues, as its bases are classified by physical scattering processes. In addition to massive HEFT, our program can generate any dark matter EFTs and EFTs of particles with any spin. The program is publicly accessible at \url{https://github.com/zizhengzhou/MassiveAmplitude}.

The paper is structured as follows. Section~\ref{sec:Basics} covers the fundamentals of spinor helicity formalism and on-shell amplitudes. In Section~\ref{sec:Constructing}, we detail a method for constructing a complete and independent set of Lorentz structure bases. To address massive EFT basis involving identical particles, Section~\ref{sec:identical} explores a comprehensible approach. Two examples illustrating how to construct massive EFT basis with this method are presented in Section~\ref{sec:example}. Section~\ref{sec:poly_to_monomial} briefly discusses how to translate amplitude bases to EFT operators. Next, Section~\ref{sec:3pt} lists all the $3$-pt massive HEFT bases. Lastly, we conclude in Section~\ref{sec:conclusion} and provide a list of $4$-pt massive HEFT operators below dimension-$8$ in Appendix~\ref{app:d8bases}.

\section{Basics of On-shell Amplitudes}
\label{sec:Basics}
The one-to-one correspondence between an EFT operator and an on-shell non-factorizable amplitude basis allows us to construct the generic on-shell massive amplitudes of massive SM fields and then map them to the operators, which are the massive HEFT bases~\cite{Ma:2019gtx,Shadmi:2018xan,Elvang:2010jv}.
We will start by briefly reviewing the general properties of on-shell scattering amplitudes. 

On-shell scattering amplitudes are determined by Lorentz symmetry, locality, and unitarity. According to Lorentz symmetry, scattering amplitudes need to be covariant under the little groups (LGs) of external legs. As such, they are typically expressed using spinor helicity variables that are charged under LGs. These spinor helicity variables are typically decomposed from the momentum matrix as ~\cite{Arkani-Hamed:2017jhn}.
\bea
    (p_i)_{\alpha \dot{\alpha}}\equiv (p_i)_\mu(\sigma^\mu)_{\alpha \dot{\alpha}}
    =\ket{i_I}_{\alpha} [i^I|_{\dot \alpha}\,,
\eea
where $\sigma^\mu \equiv \{1, \sigma^i \}$ with $ \sigma^i$ being Pauli-matrices. The right-handed spinor $\antiket{i^I}_{\dot \alpha}$ and the left-handed spinor $\ket{i^I}_{\alpha}$ correspond to the doublets of the Lorentz subgroup $SU(2)_r$ and $SU(2)_l$, respectively. This is because the Lie algebra of $SO(3,1)$ is isomorphic to $SU(2)_l \times SU(2)_r$. $I$ is the index of massive LG $SU(2)_i$. The spinors with different chirality are related by EOM, 
\bea
p_i^\mu\sigma_\mu|i^I]=m_i|i^I\rangle\,,\quad p_i^\mu\bar{\sigma}_\mu|i^I\rangle=m_i|i^I]
\eea
The massless spinors of massless momentum $p_j$, denoted as $|j]_{\dot \alpha}$ and $\ket{j}_{\alpha}$, have $\pm$ unit charge of the massless LG $U(1)_j$~\cite{Witten:2003nn}. A pair of spinors can form the minimal Lorentz invariant variables charged under LGs,
\bea
[ij]^{IJ} \equiv \epsilon^{\dot  \alpha \dot  \beta} \antiket{i^I}_{\dot\beta} \antiket{j^J}_{\dot\alpha}\,, \quad
\vev{ij}^{IJ} \equiv  \epsilon^{\alpha \beta} \ket{i^I}_{\beta} \ket{j^J}_{ \alpha}\,,
\eea 
which are called square and angle brackets. The scattering amplitude must be a function of spinor brackets. For a massive particle $i$ with spin $s_i$, its amplitude should be in the symmetric representation of the massive LG $SU(2)_i$ with rank $2s_i$. On the other hand, the amplitude of a massless particle $j$ with helicity $h_j$ must involve $2h_j$ unit charges of the massless LG $U(1)_j$.

Since the operator basis in EFT is the local interaction, the corresponding amplitude basis should be a polynomial of spinor brackets. 
Therefore, the complete set of amplitude bases consists of all the independent spinor polynomials allowed by LG symmetry, gauge symmetry, and spin statistics. We proposed a method to construct these polynomials based on the group representation theorem in ref.~\cite{Dong:2021yak, Dong:2022mcv}, which we will briefly review next.


\section{Constructing Massive Amplitude Basis}
\label{sec:Constructing}
In this section, we discuss constructing the on-shell amplitude basis for massive HEFT. To construct the basis, we use irreducible representations of the $U(N)$ and $SU(2)$ groups, i.e.~Young diagrams, to separately construct the massive LG charged part $\mathcal{B}$ and the neutral part $H$ of the amplitude basis~\cite{Dong:2021yak}. The tensor contractions of these two parts form a complete and independent set of amplitude bases $\{\mathcal{B}\cdot H\}$. However, some bases can factor out an overall mass factor, which leads to an incorrect count of their dimensions. Therefore, the second step is to minimize the dimension of the above basis~\cite{Dong:2022mcv}. To achieve this, we first construct a complete but redundant set of bases $\{\mathcal{C}\cdot F\}$ that always contains the lowest dimensional bases. We then decompose the redundant $\{\mathcal{C}\cdot F\}$ into $\{\mathcal{B}\cdot H\}$ basis from lower to higher dimensions and eliminate the linearly dependent bases. This approach helps us obtain a complete, independent and lowest-dimensional amplitude basis. We will now provide a detailed explanation of this method. 

\subsection{$\{\mathcal{B} \cdot H \}$ basis}
In general, each term of the massive amplitude basis can be decomposed into two parts: the part denoted $\mathcal{B}$ that is charged under the massive LG and the part $H$ that is neutral under the massive LG but charged under the massless LG. As the EOM can convert the LG indices of the massive left-handed spinors into right-handed spinors, we can choose the structure of $\mathcal{B}$ to be a holomorphic function of right-handed spinors. Therefore, part $\mathcal{B}$ is a tensor product of right-handed massive polarization vectors (as shown below), while part $H$ is a function of the massive momenta and massless spinors. Any amplitude basis with $m$ massive legs and $n$ massless legs can be represented as a combination of terms with the given factorization, 
\bea
\mathcal{M}^{\{I\}}_{m,n} =\sum_{\mathcal{B}\,,\, H} \sum_{\{\dot \alpha \}} \mathcal{B}^{\{I\} }_{\{\dot \alpha \} } \left(\epsilon_{i} \right) H^{\{\dot\alpha \}}\left(\antiket{j},\ket{j},p_{i} \right),
\eea
where $\epsilon_{i} \equiv \antiket{i^{\{ I_{1}}}_{\dot \alpha_1}\ldots \antiket{i^{I_{2s_i} \}}}_{\dot\alpha_{2s_i}}$ is the polarization tensor of massive particle-$i$ with spin $s_i$ and its quantum number is $(2s_i+1,2s_i+1)$ under $SU(2)_i$ and $SU(2)_r$. The bracket $\{I^i_1,\dots,I^i_{2s_i}\}$ in the $\epsilon_i$ expression denotes that these $2s_i$ indices of LG $SU(2)_i$ are entirely symmetric. We select independent polynomials, which are the results of combining the structures of $\mathcal{B}$ and $H$, as the EFT amplitude bases. By constructing independent and complete bases for $\mathcal{B}$ and $H$, we obtain the set of tensor contractions $\{\mathcal{B} \cdot H \}$, which is also independent and complete.

Since $\mathcal{B}$ is a holomorphic function of right-handed spinors, it is free from EOM and IBP redundancies. Additionally, $\mathcal{B}$ is also a linear function of polarization $\epsilon_i$, so any $\mathcal{B}$ basis must belong to the outer product of the $SU(2)_r$ representations of the $m$ massive $\epsilon_i$s.
\bea 
\mathcal{B} \subset \otimes_{i=1}^m 
    \Yvcentermath1 \renewcommand\arraystretch{0.05} \setlength\arraycolsep{0.2pt}
    \underbrace{\yng(1)\ \cdots\ \yng(1)}_{2s_i}
\eea 
The r.h.s. can be decomposed into some irreducible representations, which correspond to the $\mathcal{B}$ bases.

The $\{H\}$ basis may exhibit both EOM and IBP redundancies. To eliminate them, we can first construct massless limit bases for $H$ using $U(N)$ Semi-Standard Young Tableau (SSYT) and then map them back to the massive bases. The mapping from $H$ to its massless limit $h$ may be many-to-one, which means that different massive $H$ bases having the same massless limit should be EOM redundant. This can be seen as follows:  if $H_1$ and $H_2$ are both mapped to $h$ simultaneously, then the difference $(H_1\!-\!H_2)$ is the product of a lower-dimensional basis $H$ and a mass factor $m^2$, which denotes the EOM redundancy term. If we just choose one of them as the independent basis, the EOM redundancy can be removed during this process.  However, since $h$ contains no IBP redundancy (SSYT can remove it automatically), when it is mapped back to $H$ one-to-one (any of $H_1$ and $H_2$), this set contains neither EOM nor IBP redundancies. It should be noted that the massless limit for momentum $|i^{I}]\langle i_{I}|$ and polarization vector $|i^{\{I}]\langle i^{J\}}|$ is the same. To avoid this issue, the amplitude basis should be decomposed into $\mathcal{B}$ and $H$.

The construction process for $h$ is similar to that of the massless EFT basis, which is a singlet under Lorentz $SU(2)_l\times SU(2)_r$ group~\cite{Henning:2019enq}. In contrast, $h$ is a $SU(2)_l$ singlet but possesses the same $SU(2)_r$ representation as $\mathcal{B}$ to ensure that $\{\mathcal{B} \cdot H \}$ is a Lorentz singlet.

Now we will briefly discuss how to construct the complete set of $h$ bases without IBP redundancy. The massless spinors of $N$ external momentums  $\tilde{\lambda}_{\dot\alpha}^k \equiv \antiket{k}$ ($\lambda_{ k \alpha} \equiv  \ket{k}$) are embedded into the (anti-) fundamental representation of $U(N)$ symmetry with $k=1,\dots, N$.
So each basis of the $U(N)$ representation (i.e., an $U(N)$ SSYT) corresponds to a polynomial of massless spinors.
Conversely, this polynomial can also be written down through the SSYT according to the permutation symmetry of the $U(N)$ indices. For example, the scalar product of a right-/left-handed spinor pair can be obtained from $U(N)$ SSYT with shape $[1^2]/[1^{N-2}]$ (the $[1^2]$ is the short notation of $[1,1]$ and so is $[1^{N-2}]$).
\bea \label{eq:YD_spinorproduct}
\Yvcentermath1   \young(i,j) &=& ( \tilde{\lambda}_{ \dot\alpha}^i  \tilde{\lambda}_{ \dot\beta}^j -\tilde{\lambda}_{ \dot\alpha}^j  \tilde{\lambda}_{ \dot\beta}^i )  =  [ij]  \nonumber \\
\Yvcentermath1 \!\!\!\!\!\!N-2 \left \{ \begin{array}{c}   \blue{ \begin{Young}
    ${k_1}$  \cr
    ${k_2}$  \cr
    $\cdot$\cr
    $\cdot$\cr
 \end{Young}}
 \end{array}\right. \!\!\!
 &=& \varepsilon^{ij k_1 \ldots k_{N-2} }\lambda_{ i \alpha} \lambda_{ j \beta}
 = \frac{\vev{ i  j}}{2}\varepsilon^{ij k_1 \ldots k_{N-2} }   \,,
\eea
where $\varepsilon^{ij k_1 \ldots k_{N-2} }$ is the epslion tensor. Notice that the columns in the SSYT associated with the $U(N)$ indices of $\lambda$ are blue to distinguish them from $\tilde{\lambda}$ indices.

Generally, for a polynomial of form $\langle\circ\circ\rangle^{L/2}[\circ\circ]^{r_2}|\circ]^{r_1-r_2}_{\dot\alpha}$ (Symbol $\circ$ represents an arbitrary index), its right-handed spinors' YD of $U(N)$ group is $[r_1,r_2]$, while its left-handed spinors' YD is $[(L/2)^{N-2}]$. The irreducible representation of this polynomial of the $U(N)$ group is the outer product of them. It can be decomposed into irreducible representations via Littlewood-Richardson rules,
\bea
    h^{\{\dot\alpha \}}(\antiket{j},\ket{j},|i\rangle [i| )\!\!\!
    &=\Yvcentermath1 N-2 \left \{ \begin{array}{c}
     \\
     \\
     \\
 \end{array}  \right.\!\!\!\!\!
 \underbrace{  \blue{  \begin{array}{ccc}
    \yng(1,1)& \cdots & \yng(1,1) \\
    \vdots &  \ddots & \vdots \\
    \yng(1,1)& \cdots & \yng(1,1)
    \end{array}}  }_{L/2}  \!\otimes \,
    \renewcommand\arraystretch{0.05} \setlength\arraycolsep{0.2pt}\Yvcentermath1\begin{array}{c} \overbrace{\yng(1)\cdots\yng(1)\cdots\yng(1)}^{r_1}\\
    \underbrace{\yng(1)\cdots\yng(1)}_{r_2} {\color{white} \ \ \ \ \ \ \ \,} \end{array}\nonumber\\
    &=\Yvcentermath1 N-2 \left \{ \begin{array}{c}
     \\
     \\
     \\
 \end{array}  \right.\!\!\!\!\!
     \begin{array}{ccccc }
\blue{ \yng(1,1) }& \blue \cdots &\blue{ \yng(1,1)} \yng(1,1)&  \cdots &\yng(1,1) \\
\blue\vdots &  \blue\ddots & \blue\vdots \quad \quad & & \\
\blue{ \yng(1,1)}& \blue\cdots  & \blue{ \yng(1,1)} {\color{white} \yng(1,1)}  & &
     \end{array}
     \!\!\!\!\!\!\!\!\!\!\!\!\!\!\!\!\!\!\!\!\!\!\!\!\!\!\!\!\!\!\!\!\!\!\!\!\!\!
     \!\!\!\!\!\!\!\!\!\!\!\!\!\!
     \renewcommand\arraystretch{0.05}\begin{array}{ccccccc}
\color{white}\yng(1) &\color{white}\cdots &\color{white}\yng(2) &\color{white}\cdots &\color{white}\yng(1) &\cdots  &\yng(1)\\
\color{white}\yng(1,1,1,1,1)&  &  &  &  &  &
     \end{array}
     \!\!\!\!\!\oplus  \cdots\,,
\eea
where $\oplus \cdots$ represent other irreducible representations, and they are exactly the IBP redundancy we need to discard. All the SSYTs of the first YD constitute a set of complete $h$ bases. And the SSYTs have $(L/2+2h_j)$ fillings for massless particle-$j$ and $L/2$ fillings for massive particle-$i$.

\begin{figure}
\begin{center}
\includegraphics[width=8cm]{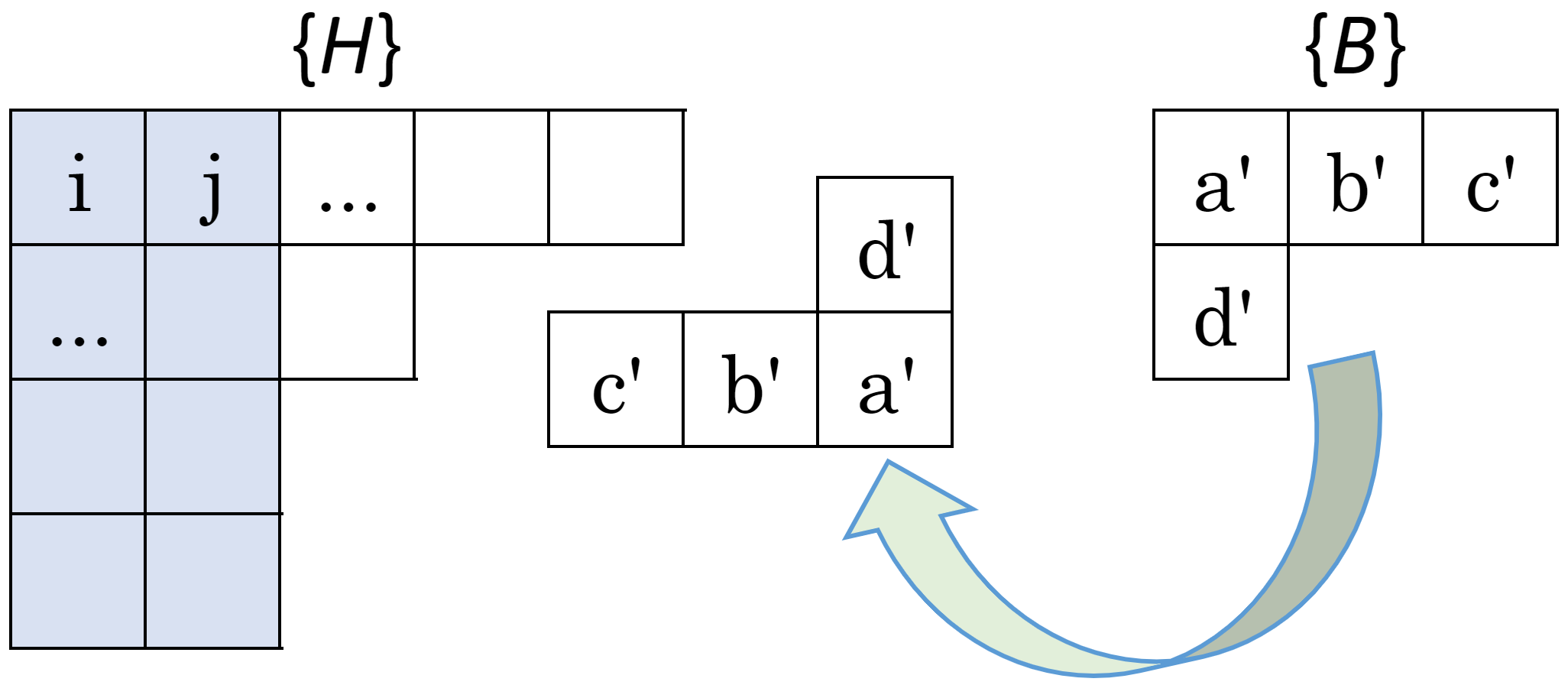}
\caption{Contracting $\mathcal{B}$ basis with $H$ basis}
\label{fig:BH_combine}
\end{center}
\end{figure}

The SSYTs of $\mathcal{B}$ and $H$ correspond only to the column permutation part of the Young operator, representing the same linear space as the Young operator. In Fig.~\ref{fig:BH_combine}, the contraction between $\mathcal{B}$ and $H$ $\dot\alpha$-indices is presented, whereby the SSYT of $\mathcal{B}$ is rotated by $180\degree$ and attached to that of $H$.
The contraction conventions of $\mathcal{B}$ and $H$ are not unique, but it was proven that the $\{\mathcal{B}\cdot H\}$ set obtained by different conventions are equivalent~\cite{Dong:2022mcv}.
It is important to note that the indexes in $\mathcal{B}$ are identified as $i^\prime$ to differentiate them from $i$ in $H$.


\subsection{$\{\mathcal{C} \cdot F \}$ basis}
\label{sec:CF}
The contraction of $\mathcal{B}$ and $H$ may contain an overall mass factor, either explicitly or implicitly. Although the presence of the mass factor $m_i$ does not render the $\mathcal{B} H$ basis redundant in the existing set, it increases the dimensionality of the set $\{\mathcal{B} \cdot H \}$ beyond its minimum. These mass factors come from the transformation between different configurations of the same polarization tensor, as exemplified by the case of massive particle-$i$,
\bea \label{eq:polarizationConfig}
\epsilon_i^{l_i} \equiv \left( |i^{\{I}\rangle \right)^{l_i}  \left(|i^{I\}}]\right)^{2s_i-l_i}\,, \quad  l_i \in [0\,,\, 2s_i]\,.
\eea
Based on the EOM, $\epsilon_i^{l_i}$ provides an overall factor of $m_i^{l_i}$. Next, we proceed to enumerate all the polarization configurations $\{l_i\}$ of the massive outlines and construct their complete bases,
\bea
\left( \mathcal{C}\cdot F \right)^{\{l_i\}}= 
 \mathcal{C}^{\{l_i\}}\left( |i^{\{I}]^{2s_i -l_i} \right)\cdot
F^{\{l_i\}}\left( |i^{I\}}\rangle^{l_i}, p_i, |j], |j\rangle \right), 
\eea
The union $\{ \mathcal{C}\cdot F \}\equiv\bigcup_{\{l_i\}}\{ \mathcal{C}\cdot F \}^{\{l_i\}}$ represents a collection of redundant but complete bases, where any lowest-dimensional amplitude basis can be found. We decompose the elements in $\{ \mathcal{C}\cdot F \}$ into the $\{ \mathcal{B}\cdot H \}$ set in ascending order of dimension and eliminate the linearly dependent terms, then we get the complete, independent and lowest-dimensional amplitude basis.

Part $\mathcal{C}^{\{l_i\}}$ is a holomorphic function of right-handed spinors $|i^I]^{2s_i -l_i}$ obtained from the polarization tensor. On the other hand, part $F^{\{l_i\}}$ is charged by both massless and massive LG due to the rest part  $|i^I]^{l_i}$. The construction of $\{ \mathcal{C}\cdot F \}^{\{l_i\}}$ is similar to that of $\{ \mathcal{B}\cdot H \}$ and is considered a special case
\bea
\{ \mathcal{B}\cdot H \}=\{ \mathcal{C}\cdot F \}^{\{l_i=0\}}
\eea
In this regard, we present a specific construction method of the amplitude basis $\{\mathcal{C}\cdot F\}^{\{l_i\}}_d$ corresponding to the $d$-dimensional EFT operator, including the shape and filling rules of the SSYT. We will discuss it later. 

By following the corresponding rules, an EFT operator can be mapped to an amplitude basis.
\bea \label{eq:matching}
&&\phi_i \leftrightarrow {\bf1} \,, \quad   \psi_{iL}, \psi_{iR}^\dagger \leftrightarrow \ket{i^I}\,, \quad \psi_{iR} , \psi_{iL}^\dagger\leftrightarrow \antiket{i^I} \,, \quad F^+_{i\mu \nu } \bar{\sigma}^{\mu \nu}  \leftrightarrow \antiket{i^{\{I_1}} \antiket{i^{I_2\}}}\,, \nonumber \\
&& m_i A_{i\mu} \sigma^\mu \leftrightarrow  \antiket{i^{\{I_1}} \langle i^{I_2\}} | \,, \quad
F^-_{i \mu\nu} \sigma^{\mu\nu} \leftrightarrow \ket{i^{\{I_1}} \ket{i^{I_2\}}} \,, \quad \partial_i \leftrightarrow  p_{i}=\antiket{i^{I}} \langle i_{I} |\,,
\eea 
where $\phi_i$ is scalar, $\psi_{iL,R}$ is a left-handed or right-handed massive fermion, $A_{i\mu} $ is a massive vector field, $F^{\pm}_{i \mu\nu} = 1/2( F_{i\mu \nu} \pm i/2 \epsilon_{\mu \nu \rho \sigma} F^{\rho \sigma}_i)$, $\sigma^{\mu \nu} \equiv \bar{\sigma}^{[\mu} \sigma^{\nu]}$, and $\bar{\sigma}^{\mu \nu} \equiv \sigma^{[\mu} \bar{\sigma}^{\nu]}$. For the massless particle, the mappings are similar, except that there is no bare massless gauge field $A_\mu$ in amplitude basis. According to these mappings, the dimension of EFT operators is,
\bea \label{eq:dmension} 
d=[\mathcal{M}] +N-n_A\,,
\eea
where $[\mathcal{M}]$ is the energy dimension of the amplitude basis, and $n_A$ is the number of bare vector fields $A_\mu$. So the set $\{\mathcal{C}\cdot F\}^{\{l_i\}}_d$ can be constructed by the following procedures:  
\begin{itemize}
    \item[$\bullet$] Fill in $(L/2-l_i)$ number-$i$, $(2s_i-l_i)$ number-$i'$ and $(L/2+2h_j)$ number-$j$ in the glued Young diagram of shape $[(L+R)/2\,,\,  (L+R)/2\,,\,  \left(L/2 \right)^{N-4}]$, where
    \bea\label{eq:YCF}
    R &=&d-N+n_A +\sum (s_i -l_i)+\sum h_j \,, \nonumber \\
    L &=&d-N+n_A -\sum (s_i -l_i) -\sum h_j\,.
    \eea    
    
    \item[$\bullet$] Number-$i'$ can only be filled in the $L/2$ white columns because it comes from $\mathcal{C}^{\{l_i\}}$.
    
    \item[$\bullet$] Massless limit $\rightarrow$ amplitude basis: map any $l_i$ left-handed spinor $\ket{i}$s and $(2s_i -l_i)$ right-handed spinor $|i^\prime]$s to $\epsilon_i^{l_i}$; map the remaining $\ket{i}$ and $|i]$ pairs to $|i^I] \langle i_I|$.
\end{itemize}

Next we decompose $\{\mathcal{C}\cdot F\}$ to $\{\mathcal{B}\cdot H \}$ from low to high dimension and eliminate the linearly dependent bases. 
Since the difference between different $\mathcal{B}H$s with the same massless limit is EOM redundancy, we only need to decompose $\{\mathcal{C}\cdot f\}$ to $\{\mathcal{B}\cdot h \}$ in the massless limit to obtain the independent lowest-dimensional basis~\cite{Dong:2022mcv}. For the dimension-$d$ set $\{\mathcal{B}\cdot H\}_d$, we have
\bea \label{eq:Dmension}
\{\mathcal{B}\cdot h\}_d\sim\bigcup_{\{l_i\}}\{\mathcal{C}\cdot f\}_{d-\sum l_i-n_A}
\eea
Note that the decomposition process of $\{\mathcal{B}\cdot h\}_d$ for different $d$ is independent. The following are the step-by-step instructions for the decomposition process:
\begin{itemize}
    \item[1.] Input the amplitude bases in $\{\mathcal{C}\cdot f\}$ from low to high according to $(d-\sum l_i-n_A)$, and multiply each of them by the corresponding factor $\prod_i[ii']^{l_i}\,$;

    \item[2.] Replace all occurrences of $|1]\bra{1}$ using equation $\sum_{i=1}^N|i]\bra{i}=0\,$;

    \item[3.] Replace all occurrences of $[23]\vev{32}$ using equation $\left(\sum_{i=2}^N|i]\bra{i}\right)^2=0\,$;

    \item[4.] Apply Schouten identities to transform the Young tableaux corresponding to left- and right-handed spinors into semi-standard forms separately, and the number order is $ 1<\cdots<N<N'<\cdots<1'\,$;
    
    \item[5.] Repeat steps 3 and 4 until no $[23]\vev{32}$ appears. Now we represent this $(\mathcal{C}\cdot f)$ basis as a vector in linear space $\{\mathcal{B}\cdot h\}_d$.
\end{itemize}
Note that when marking the particle number, the first particle should be kept massive; otherwise, the above process will become more cumbersome. 
The coefficients before each $(\mathcal{B}\cdot h)$ basis form a coefficient vector, then we arrange the vectors that represent $(\mathcal{C}\cdot f)$ from low to high in $(d-\sum l_i-n_A)$ and eliminate terms that are linearly dependent. Afterward, the remaining vectors give the lowest-dimensional amplitude basis which is equivalent to the set $\{\mathcal{B}\cdot H\}_d$. By traversing all possible values of $d$ in $\{\mathcal{B}\cdot H\}$, we obtain the amplitude bases corresponding to the general EFT operators, which we denote as $\{\mathcal{L}\}$.
              

\section{Basis Involving Identical Particles}
\label{sec:identical}

The previous section addressed the construction of basis for distinguishable particles. In this section, we discuss constructing basis involving identical particles. The amplitude basis for $n$ identical bosons (fermions) must be in the totally (anti-)symmetric representation of the permutation group $S_n$, corresponding to the Young diagram $[n]$ ($[1^n]$). 
To achieve this, the Young operator $\mathcal{Y}_{[R_n]}$ of the $S_n$ representation $[R_{n}] = [n]$ or $[1^n]$ is used, which makes the wave function of the $n$ identical particles totally symmetric or anti-symmetric. The eigenstates of the $\mathcal{Y}_{[R_n]}$ operation must be in the $[R_n]$ representation. To construct a complete amplitude base in the $[R_n]$ representation, we first use the Young operator $\mathcal{Y}_{[R_n]}$ to act on the space of the complete basis $\{\mathcal{M}\}_d$ at dim-$d$. We then obtain the representation matrix $M_{[R_n]}$ of $\mathcal{Y}_{[R_n]}$. Finally, the eigenvectors with non-zero eigenvalues (actually is $1$) correspond to the amplitude bases in the $[R_{n}]$ representation, whereas eigenvectors with zero eigenvalues correspond to bases that vanish under identical particle permutations.

In the previous section, we discussed how to construct independent Lorentz structures. Generally, an amplitude basis is composed of two parts: the gauge structure ($T$) and the Lorentz structure ($\mathcal{L}$),
\bea
    \mathcal{M} = T \times \mathcal{L}\,.
 \eea
The complete set of amplitude bases can be constructed by multiplying the complete gauge structure bases and Lorentz structure bases,
\bea
\{ \mathcal{M} \} = \{T\} \times \{\mathcal{L}\}\,.
\eea
It should be noted that any Young operator of the permutation group $S_n$ can be expressed as a function of the permutation elements $(12)$ and $(1\ldots n)$. Therefore, it is sufficient to obtain the representation matrices $M^T_{(12),(1,\ldots,n)}$ and $M^{\mathcal{L}}_{(12),(1,\ldots,n)}$ in the $\{T\}$ and $\{\mathcal{L}\}$ space respectively, for the required Young operator of $S_n$. Consequently, the representation matrix $M_{[R_n]}$ is determined by computing the outer product of the matrices $M^T$ and $M^{\mathcal{L}}$. For example, the matrix of totally symmetric representation $[3]$ of the $S_3$ group can be expressed as
\eq{ \label{eq:B/Fmatrix}
    &M_{\scriptsize\young(123)}=(\mathds{1}+x+y+xy+yx+xyx)/6\,,\\
    &x=M^T_{(12)}\otimes M^{\mathcal{L}}_{(12)},\quad y=M^T_{(123)}\otimes M^{\mathcal{L}}_{(123)},
}
Next, we will introduce how to construct the gauge structure basis and obtain the representation matrices in the $\{T\}$ and $\{\mathcal{L}\}$ space.
\subsection{Gauge Structure}
This subsection describes the construction of the gauge structure basis $T$ and the $M^{T}$ matrix, responsible for identical permutation representation, under the $SU(3)_c$ color symmetry. The same method can be applied to the $SU(N)$ gauge group. For the fundamental representation of $SU(3)_c$, the Standard Young Tableaux (SYTs) of up quark, anti-up quark, and gluons are
\eq{ \label{eq:color_inside_symmetry}
    u^{c_1}\sim\Yvcentermath1\young(\onec)
    \quad\quad\quad
    \epsilon^{c_x c_1 c_2 }\,\bar{u}_{c_x}\sim\Yvcentermath1\young(\onec,\twoc)
    \quad\quad\quad
    \epsilon^{c_x c_1 c_2}\,(\lambda_i)_{c_x}^{c_3}\,g^i\sim\Yvcentermath1\young(\onec \threec,\twoc)
}
where $\lambda^{i}$ refers to the eight Gell-Mann matrices. The set $\{T\}$ of color structures represents all $SU(3)_c$ singlets in the outer product of SYT of the relevant particles. Instead of selecting singlets from the outer product, a more efficient method is used to construct $\{T\}$. Firstly, all the singlet SYTs in the fundamental representation are listed. From these, the components that satisfy the symmetry given by the SYT of each relevant particle are projected as our result.

Consider the four-point scattering process $u\!-\!\bar{u}\!-\!g_1\!-\!g_2$ as an example. To construct the basis $\{T\}$, we start by listing all SYTs in the shape of $[3\,, 3\,, 3]$ that satisfy the $SU(3)_c$ index permutation symmetries of $\bar{u}$ as well as $g^i_{1,2}$. There are exactly 42 such SYTs and we denote this list as $\{T^\prime\}$. To select the $\{T\}$ basis from $\{T^\prime\}$, we use the Young operators of the external legs' SYTs to act on the $\{T^\prime\}$ SYTs, which yields a $42 \times 42$ matrix $P$,
\bea \label{eq:numberbox}
\mathcal{Y}_{\scriptsize\young(\onec)} \times \mathcal{Y}_{\scriptsize\young(\twoc,\threec)} \times \mathcal{Y}_{\scriptsize\young(\fourc \fivec,\sixc)}  \times \mathcal{Y}_{\scriptsize\young(\sevenc \eightc,\ninec)} \, \{T^\prime \} \equiv \mathcal{P} \cdot \{T^\prime \} \,. 
\eea
where the $SU(3)_c$ index of each of the four particles is marked by the color indices $1^c,\ldots,9^c$. The SYTs in $\{T^\prime\}$ are not standard after being acted on by the Young operators. We can obtain the coefficient matrix $P$ by decomposing the symmetrized $T^\prime$ back into the linear space of $\{T^\prime\}$. The eigenvectors of $P$ with an eigenvalue of 1 correspond to the $\{T\}$ set required. As the rank of the $P$ matrix is three, there exist three $T$ bases,
\begin{equation} \label{eq:Tbases}
    \{\mathrm{T}\}= \mathcal{Y}_{\scriptsize\young(\onec)} \times \mathcal{Y}_{\scriptsize\young(\twoc,\threec)} \times \mathcal{Y}_{\scriptsize\young(\fourc \fivec,\sixc)}  \times \mathcal{Y}_{\scriptsize\young(\sevenc \eightc,\ninec)} \left\{
    \Yvcentermath1
    \young(\onec \twoc \threec,\fourc \fivec \sixc,\sevenc \eightc \ninec) \  ,\
    \young(\onec \twoc \threec,\fourc \fivec \sevenc,\sixc \eightc \ninec) \ ,\
    \young(\onec \twoc \fourc,\threec \sixc \sevenc,\fivec \eightc \ninec)\right\}\,,
\end{equation}
where $T$ basis is obtained by multiplying the $T^\prime$ basis with the Young operator. When reading the amplitude basis from the $T$ basis, the Young operator should be read as the tensor factor in front of the field operator in eq.~(\ref{eq:color_inside_symmetry}), while $T^\prime$ should be read as total antisymmetric tensors. For instance,
\bea \label{eq:Tstructure}
\Yvcentermath1 \young(\onec \twoc \threec,\fourc \fivec \sixc,\sevenc \eightc \ninec)  \sim \epsilon^{c_1 c_4 c_7} \epsilon^{c_2 c_5 c_8}  \epsilon^{c_3 c_6 c_9} \,. 
\eea

After obtaining the $\{T\}$ basis, the matrix $M^T_{(12),(1,\ldots,n)}$ can be obtained by acting the permutation element on the color indices of the identical particles. For example, for identical gluons $g_1$ and $g_2$, the permutation element $(12)$ exchanges the color indices $4^c\leftrightarrow 7^c,\,5^c\leftrightarrow 8^c$, and $6^c\leftrightarrow 9^c$ in the SYTs. On decomposing the resultant non-standard Young tableaux into combinations of the SYTs, we obtain the matrix $M^{T}_{(12),(1,\ldots,n)}$. 
For instance, by applying the element $(12)$ to the color indices of $\{T^\prime\}$ SYTs and breaking down the resulting Young tableaux into SYTs, we obtain the matrix $M^{T^\prime}_{(12)}$ in the $\{T^\prime \}$ space. This matrix can be expressed as
\bea
    M^T_{(12)}=M^{T^\prime}_{(12)}\cdot P\,, 
\eea
where matrix $P$ satisfies $P^2=P$ due to the property that its eigenvalues are 1 and 0. In addition, $[P,M^{T^\prime}]=0$ because the Young operators commute with both elements $(12)$ and $(1,\ldots,n)$.
Shifting our focus to eq.~(\ref{eq:B/Fmatrix}), we proceed to commute the $P$ matrix to the rightmost position. This operation transforms the representation of $S_3$ into a totally symmetric one, yielding the following result.
\eq{ \label{eq:S3_symmetrize}
    &M_{\scriptsize\young(123)}=(\mathds{1}+X+Y+XY+YX+XYX) (P\otimes\mathds{1}^{\mathcal{L}})/6,\\
    &X=M^{T^\prime}_{(12)}\otimes M^\mathcal{L}_{(12)},\quad Y=M^{T^\prime}_{(123)}\otimes M^\mathcal{L}_{(123)}\,,}
where $\mathds{1}^{\mathcal{L}}$ is the identical matrix in the linear space of Lorentz structures.

\subsection{Lorentz Structure} 
\label{sec:Lorentz_structure}
In order to construct the $M^\mathcal{L}$ matrix in the space of the complete Lorentz basis $\{ \mathcal{L} \}$ with the lowest dimension, we employ a similar procedure to that used for constructing the $M^T$ matrix. Specifically, we apply the permutation $(12)$ to the elements in $\{\mathcal{L}\}$ and then decompose the resulting bases into combinations of elements from $\{\mathcal{L}\}$. It is worth noting that since the permutation $(12)$ leaves the operator dimension, $n_A$, and $\sum l_i$ unchanged, so the basis that we obtain through this decomposition satisfy
\eq{ \label{eq:Dd_lower}
    \bigcup_{D'\leq D,d'\leq d}\{\mathcal{L}\}^{D'}_{d'}
    \ \leftarrow\ 
    (12)\{\mathcal{L}\}^D_d\,,
}
where $D\equiv d+n_A+\sum l_i$. Since the set $\{\mathcal{L}\}$ is of the lowest dimension, $d'\leq d$, and the completeness of the $\{\mathcal{B}\cdot H\}$ set\footnote{By retaining the LG indices during the $\{\mathcal{B}\cdot H\}$-decomposition (Section \ref{sec:CF}), an arbitrary dim-$d$ amplitude basis can be decomposed into the set $\bigcup_{x\leq d+n_A+\sum l_i}\{\mathcal{B}\cdot H\}_x$.} leads to $D'\leq D$. This implies that both the representation matrices, $M^{\mathcal{L}}_{(12)}$ and $M_{[n]}$, as in Equation (\ref{eq:S3_symmetrize}), are block lower triangular matrices ordered by $D$ and $d$ in the linear space. Moreover, the idempotent nature of the Young operator $\mathcal{Y}_{[n]}$ implies that $M_{[n]}$ is an idempotent matrix with eigenvalues of 1 and 0, and so does $M_{[1^n]}$. 

According to linear algebra, the eigenvectors of an lower triangular idempotent matrix with an eigenvalue of 1 are determined by the diagonal matrix. The diagonal matrix $(M_{(12)}^{\mathcal{L}})^D_d$ in the subset $\{\mathcal{L}\}^D_d$ can be obtained by first using the $\{\mathcal{B}\cdot h\}$-decomposition to obtain $(M_{(12)}^{\mathcal{L}})^D$ and then extracting the diagonal blocks of different $d$. To obtain matrix $(M_{(12)}^{\mathcal{L}})^D$, we decompose the bases in set $(12)\{\mathcal{L}\}^D$ back to the set $\{\mathcal{L}\}^D$ in the massless limit. We use $\{\mathcal{B}\cdot h\}_D$ as an auxiliary intermediate set for mapping purposes as
\bea
    \{\mathcal{L}\}^D \leftarrow \{\mathcal{B}\cdot h\}_D \leftarrow (12)\{\mathcal{L}\}^D\,,
\eea
and the representation matrix is
\bea
(M^{\mathcal{L}}_{(12)})^D=[M^{\mathcal{B}h\leftarrow  \mathcal{L}}_{e}]^{-1}\cdot M^{\mathcal{B}h\leftarrow  \mathcal{L}}_{(12)}\,.
\eea 
where $M^{\mathcal{B}h\leftarrow  \mathcal{L}}_{(12)}$ is the coefficient matrix from decomposing the elements in $(12)\{\mathcal{L}\}^D$ into $\{\mathcal{B}\cdot h\}_D$, and $M^{\mathcal{B}h\leftarrow  \mathcal{L}}_{e}$ is similar.


\section{Some Examples} 
\label{sec:example}
In this section, we provide examples to illustrate the construction of the basis involving identical particles.

\subsection{Basis of $Z\!-\!Z\!-\!Z\!-\!h$}
In the following, we outline the construction of the amplitude basis of $ZZZh$ at dim-$5$. This basis should belong to the representation $[3]$ of $S_3$. Referring to Eq.~(\ref{eq:Dmension}), we identify a total of nine dim-$5$ Lorentz bases involving three distinguishable massive vectors $V_{1,2,3}$ and one scalar $s$. Specifically, there are three bases at $D=9$ and six at $D=11$,
\bea \label{eq:ZZZhbasis}
\renewcommand{\arraystretch}{1.2}
   \{ \mathcal{L} \}^{\fd=9}_{\pd=5} (V_1 V_2V_3 s) =\!\left \{
    \begin{array}{l}
        \abk{12}\sbk{1^{\prime}3^{\prime}}\sbk{2^{\prime}3^{\prime}} \\
        \abk{13}\sbk{1^{\prime}2^{\prime}}\sbk{2^{\prime}3^{\prime}} \\
        \abk{23}\sbk{1^{\prime}2^{\prime}}\sbk{1^{\prime}3^{\prime}} \\
    \end{array}
    \right \}\,,
   \{ \mathcal{L} \}^{\fd=11}_{\pd=5} (V_1 V_2V_3 s)=\!\left \{
    \begin{array}{l}
        \abk{12}\abk{13}\sbk{2^{\prime}3^{\prime}}                  \\
        \abk{12}\abk{23}\sbk{1^{\prime}3^{\prime}}                  \\
        \abk{13}\abk{23}\sbk{1^{\prime}2^{\prime}}                  \\
        \abk{13}\abk{23}\sbk{32^{\prime}}\sbk{1^{\prime}3^{\prime}} \\
        \abk{13}\abk{24}\sbk{42^{\prime}}\sbk{1^{\prime}3^{\prime}} \\
        \abk{12}\abk{34}\sbk{43^{\prime}}\sbk{1^{\prime}2^{\prime}} \\
    \end{array}
    \right \}.
\eea
Note that we employ the temporary massless limit to express the amplitude bases.
Since both $Z$ and $h$ are $SU(3)_c$ singlets, their $T$ bases become trivial, resulting in $\mathcal{M}=\mathcal{L}$. Hence, we only need to compute the matrices $M_{(12)}^{\mathcal{L}}$ and $M_{(123)}^{\mathcal{L}}$. By extracting the diagonal block matrices within the $D=9$ and $11$ basis, we can derive the amplitude bases in the $S_3$ representation $[3]$, which are     
\begin{equation}
    \renewcommand{\arraystretch}{1.4}
    (M_{(12)} ) ^{\fd=9}_{\pd=5} =
    \setlength{\arraycolsep}{3.3pt}
    \begin{pmatrix}
        -1 & 0  & 0  \\
        0  & 0  & -1 \\
        0  & -1 & 0  \\
    \end{pmatrix}\, ,\quad \quad  
    \renewcommand{\arraystretch}{1.4}
    (M_{(123)})^{\fd=9}_{\pd=5} =
    \setlength{\arraycolsep}{7pt}
    \begin{pmatrix}
        0 & 0 & 1 \\
        1 & 0 & 0 \\
        0 & 1 & 0 \\
    \end{pmatrix}\,.
\end{equation}

\begin{equation}
    (M_{(12)} ) ^{\fd=11}_{\pd=5} =
    \renewcommand{\arraystretch}{1.4}
    \setlength{\arraycolsep}{4pt}
    \begin{pmatrix}
        0  & -1 & 0  & 0  & 0  & 0 \\
        -1 & 0  & 0  & 0  & 0  & 0 \\
        0  & 0  & -1  & 0  & 0  & 0 \\
        0  & 0  & 1  & 1  & 0 & 0 \\
        -1 & 1  & -1 & -2 & -1 & 1 \\
        0  & 0  & 0  & 0  & 0  & 1 \\
    \end{pmatrix}\,, \quad 
     (M_{(123)})^{\fd=11}_{\pd=5} =
    \begin{pmatrix}
        0  & 1 & 0  & 0  & 0  & 0 \\
        0  & 0 & 1  & 0  & 0  & 0 \\
        1  & 0 & 0  & 0  & 0  & 0 \\
        -1 & 0 & 0 & -1 & -1  & 0 \\
        0  & 0 & 0  & 0  & 0  & 1 \\
        -1 & 1 & -1 & -2 & -1 & 1 \\
    \end{pmatrix}\,.
\end{equation}
According to the expression of the Young operator matrix of $[3]$ in Eq.~(\ref{eq:B/Fmatrix}),  we can get the diagonal block matrices of  $M_{\tiny\young(123)}$,  
\begin{align}
    (M_{\tiny\young(123)} )^{D=9}_{d=5} = 0 \,, \quad
   ( M_{\tiny\young(123)} )^{D=11}_{d=5} = \frac{1}{6}
   \renewcommand{\arraystretch}{1.4}
    \setlength{\arraycolsep}{4pt}
    \begin{pmatrix}
        0  & 0  & 0  & 0 & 0  & 0  \\
        0  & 0  & 0  & 0 & 0  & 0  \\
        0  & 0  & 0  & 0 & 0  & 0  \\
        1  & -1 & 1  & 2 & 0  & -2 \\
        -2 & 2  & -2 & -4 & 0 & 4  \\
        -2 & 2  & -2 & -4 & 0 & 4  \\
    \end{pmatrix}\,.
\end{align}
The two projection matrices above illustrate that only one dim-5 operator complies with the spin statistics in the $\{\mathcal{L}\}^{\fd=11}_{\pd=5} (V_1 V_2V_3 s)$ subset. 
The eigenvector, with the eigenvalue 1, can be represented as $M_{\tiny\young(123)} \cdot (e_4)^{D=11}_{d=5}$, where $(e_i)^D_d$ symbolizes the $i$-th basis in the $\{\mathcal{L}\}^D_d$ set. 
Using the Young operator $\mathcal{Y}_{\tiny\young(123)}$ to symmetrize the amplitude basis produces the only dim-5 operator of scattering $Z^3h$. 
\bea \label{eq:fullZZZh}
\mathcal{M}_{d=5} (Z^3h) 
&=&\mathcal{Y}_{\tiny\young(123)}\abk{\bf 13} \abk{\bf 23}\sbk{\bf 23}\sbk{\bf 13} ,
\eea
where the basis $\{e_4\}^{\fd=11}_{\pd=5} = \abk{13}\abk{23}\sbk{2^\prime 3}\sbk{1^{\prime}3^{\prime}}$ is mapped into the massive one based on the rules outlined in Sec.~\ref{sec:CF}. The bolding spinor $|\bf i]$ and $\ket{\bf i}$ are employed to represent the massive spinors for simplicity.

\subsection{Basis of $u\!-\!\bar{u}\!-\!g\!-\!g$}
The gauge structure basis of $u\bar{u}gg$ has nontrivial bases for up quarks and two gluons both with $+1$ helicity, as shown in Eq.~(\ref{eq:Tbases}). Using the technique outlined in Sec.~\ref{sec:Constructing}, the Lorentz structure basis for fermion ($f$), antifermion ($\bar{f}$), and two massless vectors ($v_1$ and $v_2$) at dim-8 can be constructed. We find that there exist a total of two Lorentz bases.
\begin{equation}
\renewcommand{\arraystretch}{1.2}
    \{\mathcal{L} \} ^{\fd=9}_{\pd=8} (f \bar{f} v_1^+ v_2^+)= \left\{
    \begin{array}{l}
        \abk{13}\sbk{34}\sbk{34}\sbk{32^{\prime}} \\
        \abk{23}\sbk{34}\sbk{34}\sbk{31^{\prime}} \\
    \end{array}
     \right\} \,.
 \end{equation}   
Multiplying bases in $\{\mathcal{L}\}$ with bases in $\{T\}$ produces six bases for a quark, an antiquark, and two distinguishable gluons.  
\bea \label{eq:Mbases}
\{\mathcal{M} \}^{\fd=9}_{\pd=8} (f \bar{f} v_1^+ v_2^+) = \mathcal{P}\cdot\left\{
    \Yvcentermath1
    \young(\onec \twoc \threec,\fourc \fivec \sixc,\sevenc \eightc \ninec) \  ,\
    \young(\onec \twoc \threec,\fourc \fivec \sevenc,\sixc \eightc \ninec) \ ,\
    \young(\onec \twoc \fourc,\threec \sixc \sevenc,\fivec \eightc \ninec)\right\}\otimes \{\mathcal{L} \} ^{\fd=9}_{\pd=8} (f \bar{f} v_1^+ v_2^+)\,.
\eea
Assuming $v_1^+$ and $v_2^+$ represent identical gluons, their bases should fall under the $S_2$ representation $[2]$. Consequently, we can obtain the Lorentz and gauge structure matrix $M_{(34)}^{\mathcal{L}}$ and $M_{(34)}^{T}$ for $S_2$ element $(34)$,
\bea
    \renewcommand{\arraystretch}{1.4}
    \setlength{\arraycolsep}{2.8pt}
   (M_{(34)}^{\mathcal{L}} )^{\fd=9}_{\pd =8} =
    \begin{pmatrix}
        -1 & 0  \\
        0  & -1 \\
    \end{pmatrix}\, ,  \quad  (M^{T}_{(34)}) ^{\fd=9}_{\pd =8}=
    \begin{pmatrix}
        14/3  & 22/3 & -2 \\
        -11/6 & -8/3 & 1  \\
        11/3  & 22/3 & -1 \\
    \end{pmatrix}.
 \eea   
Then we can get the matrix $M_{\tiny\young(34)}$ of  $\mathcal{Y}_{\tiny\young(34)}$ in the linear space of the amplitude bases, 
\bea
    \renewcommand{\arraystretch}{1.4}
    \setlength{\arraycolsep}{2.8pt}
    M_{\text{\tiny \young(34)}} =
    \frac{1}{2}\left(1+ (M^{T}_{(34)})\otimes  (M_{(34)}^{\mathcal{L}} ) \right) = \frac{1}{12}\begin{pmatrix}
        -22 & 0   & -44 & 0   & 12 & 0  \\
        0   & -22 & 0   & -44 & 0  & 12 \\
        11  & 0   & 22  & 0   & -6 & 0  \\
        0   & 11  & 0   & 22  & 0  & -6  \\
        -22 & 0   & -44 & 0   & 12 & 0  \\
        0   & -22 & 0   & -44 & 0  & 12 \\
    \end{pmatrix}\,.
\eea 
We observe that there exist two eigenvectors with nonzero eigenvalues: $M_{\text{\tiny \young(34)}} \cdot (e_1)^{\fd=9}_{d=8}$ and $M_{\text{\tiny \young(34)}} \cdot (e_2)^{\fd=9}_{d=8}$, where $(e_{1,2})^{\fd =9}_{d=8}$ correspond to the first and second amplitude bases in Eq.~(\ref{eq:Mbases}). Similar to the previous example, the complete expressions for the two amplitude bases are given by:    
\bea \label{eq:uugg}
\{\mathcal{M}\}_{d=8}^{D=11} (u\bar{u}g^+g^+) =\left \{\mathcal{P}\cdot \mathcal{Y}_{\tiny\young(34)} \; \Yvcentermath1 \young(\onec \twoc \threec,\fourc \fivec \sixc,\sevenc \eightc \ninec)  \,\abk{{\bf 1}3}\sbk{3 {\bf 2 }} \sbk{34}^2 \,, \, \mathcal{P}\cdot\mathcal{Y}_{\tiny\young(34)} \;  \Yvcentermath1 \young(\onec \twoc \threec,\fourc \fivec \sixc,\sevenc \eightc \ninec)  \,\abk{{\bf 2 } 3} \sbk{3 {\bf 1}} \sbk{34}^2 \right \} \,, 
\nonumber\\
\eea
where the Young operator $\mathcal{Y}_{\tiny\young(34)}$ acts on all the gauge indices and spinors associated with leg-3 and 4.

Likewise, constructing amplitude basis that comprise distinct types of identical bosons (fermions) simply requires multiplying the matrix $M_{[R_n]}$ for each identical particle type to obtain a total matrix. The eigenvectors with eigenvalue-1 represent the basis that satisfies the identical particle statistics.

\section{Amplitude Bases to EFT Operators}\label{sec:poly_to_monomial}
When the amplitude basis gets mapped to the EFT operator, it inherently adheres to spin statistics. The mapping process itself resembles the function of Young operator $\mathcal{Y}_{[R_n]}$. Therefore, the mapping of $\mathcal{M}_i$ is identical to mapping $\mathcal{Y}_{[R_n]}\cdot\mathcal{M}_i$ to operators as granted by Feynman rules. For instance, we can map the monomial basis (\ref{eq:fullZZZh}) for $Z^3h$ to the dim-5 EFT operator,
 \bea 
 \abk{\bf 13} \abk{\bf 23}\sbk{\bf 23}\sbk{\bf 13}   \to 
 {Z}_{ \nu} {Z}_{ \rho} \left( D_{\mu} {Z}_{ \sigma} \right) h \operatorname{Tr} \left( \sigma^{\nu} \bar{\sigma}^{\mu} \sigma^{\rho} \bar{\sigma}^{\sigma} \right).
 \eea          
Based on the QCD indices and SYTs correspondence in Eq.~(\ref{eq:color_inside_symmetry}) and the gauge structure in Eq.~(\ref{eq:Tstructure}), the first operator of $u\bar{u}g^+g^+$ in Eq.~(\ref{eq:uugg}) is:
\bea \label{eq:uugg_oprator}
&&\left( \mathcal{Y}_{\scriptsize\young(\onec)} \times \mathcal{Y}_{\scriptsize\young(\twoc,\threec)} \times \mathcal{Y}_{\scriptsize\young(\fourc \fivec,\sixc)}  \times \mathcal{Y}_{\scriptsize\young(\sevenc \eightc,\ninec)}\,\Yvcentermath1 \young(\onec \twoc \threec,\fourc \fivec \sixc,\sevenc \eightc \ninec) \right) \,\abk{{\bf 1}3}\sbk{3 {\bf 2 }} \sbk{34}^2 \to  \nonumber \\
&&(\epsilon_{c_1 c_4 c_7} \epsilon_{c_2 c_5 c_8}  \epsilon_{c_3 c_6 c_9} )  \left(\epsilon^{c_x c_2 c_3} {\bar{u}}_{L c_x} \bar{\sigma}^{\mu} {u}_{L}^{c_1} \right) \left(\epsilon^{c_y c_4 c_6} D_{\mu}   ({G}_{ \sigma \rho}^+ )_{c_y}^{c_5} \right) \left(\epsilon^{c_z c_7 c_9} (G^{+ \sigma \rho} )_{c_z}^{c_8} \right)\,,
\eea    
where $G^\pm_{\mu\nu} =1/2 (G_{\mu\nu} \pm i/2 \epsilon_{\mu\nu \rho \sigma} G^{\rho \sigma} )$, $(G_{\mu\nu})^{c_5}_{c_y} \equiv  (\lambda^{i} G_{\mu \nu}^i ) ^{c_5}_{c_y}$, and $G_{\mu \nu}^i$  is the field strength of gluon.

\section{HEFT Operators}
\label{sec:3pt}
In the SM, all particles are massless, including the Higgs doublet (H), gauge bosons of $U(1)_Y \times SU(2)_L \times SU(3)_c$ $\{ B_\mu,W_\mu^a,g_\mu^i \}$, and fermions $\{E_L,e_R,Q_L,u_R,d_R\}$, until the EWSB takes place, causing the fields to become massive except photons and gluons. 
In HEFT, the EWSB is nonlinearly realized, hence constructing HEFT basis directly in terms of SM mass eigenstates such as $\{h, \gamma_\mu,Z_\mu,W^\pm_\mu,g_\mu^i\}$ and $\{e,\nu,u,d\}$ is more practical and convenient for carrying out phenomenological research. 
In this section, we present the three-point HEFT basis for these SM fields, which coincide with ~\cite{Durieux:2019eor}. And four-point basis up to dim-8, constructed using Mathematica codes, are listed in App.~\ref{app:d8bases}.
It is noteworthy that although we only generated the EFT basis for one generation of fermions, the code is capable of creating basis for three generations as well.

As the number of 3-pt amplitude bases is finite, we can construct all of their standard Young tableaux and derive their expressions, as the construction of $\BH$ bases. The expressions for these bases can be simplified easily due to the trivial kinematics of 3-point amplitudes, allowing us to find bases with the lowest dimension. This section comprises two tables that exhibit all 3-pt amplitude bases and their associated HEFT operators. For each vertex, the number of bases is presented as $N_S$, and we have assumed the neutrino to be a Dirac fermion.
\begin{table}[H]
    \begin{tabular}{l|l|l|l}
    \hline
     \textbf{Vertex} & $ \, {\bf N_S} \,$ &  \quad \quad \textbf{Amplitude}  & \textbf{Operator} \\ \hline
    $f\bar{f}h$\,
    &\,2
    &$[{\bf 12}],\,\langle {\bf 12}\rangle$ & $h \bar{f}_R f_L$\,, \,$h \bar{f}_L f_R$ 
    \\ \hline
    $f\bar{f}\gamma^\pm$
    &\,1
    &$[{\bf 1}3][{\bf 2}3]\,/\,\langle {\bf 1}3\rangle\langle {\bf 2}3\rangle$ & $\gamma_{\mu \nu}^\pm \bar{f}_R \sigma^{\mu \nu} f_L$ /$\gamma_{\mu \nu}^\pm  \bar{f}_L \sigma^{\mu \nu} f_R$
    \\ \hline
     $ q\bar{q}g^\pm$
    &\,1
    &$[{\bf 1}3][{\bf 2}3]\,/\,\langle {\bf 1}3\rangle\langle {\bf 2}3\rangle$ & $G_{\mu \nu}^\pm \bar{q}_{R} \sigma^{\mu \nu} q_{L} \,/\, G_{\mu \nu}^\pm \bar{q}_{L} \sigma^{\mu \nu} q_{R}$  
    \\ \hline
    $f\bar{f} V $\,  
    &\,4
    &$[\bf 13]\langle23\rangle,\,[23]\langle13\rangle,\,[13][23],\,\langle13\rangle\langle23\rangle$   &  $\bar{f}_L \slashed{V} f_R$,  $\bar{f}_R \slashed{V} f_L$, $V^+_{\mu \nu} \bar{f}_L \sigma^{\mu \nu} f_R$, $V^-_{\mu \nu} \bar{f}_L \sigma^{\mu \nu} f_R$ \\ \hline
    \end{tabular}
    \caption{The $3$-pt bases for fermions, where $f=\{u,d,e,\nu\}$, $q=\{u,d\}$ , and $V=\{Z, W^\pm \}$.}
\end{table}
The above table enumerates all amplitudes comprising SM fermions, which are simpler since they have no identical particles. Subsequently, we present the 3-pt EFT operators without fermions.
\begin{table}[H]
\begin{tabular}{l|l|l|l }
\hline
\textbf{Vertex} & $ {\bf N_S} $   & \quad \quad \quad \textbf{Amplitude}  & \quad \quad \textbf{Operator} \\ \hline
$hhh$
&1
&$Const.$
&$hhh$
\\ \hline

$h\gamma^\pm \gamma^\pm \, (hg^\pm g^\pm)$
&1
&$[23][23]\,/\,\langle23\rangle\langle2{3}\rangle$
&$h\gamma_{\mu\nu}^{\pm}\gamma^{\pm \mu\nu}\, (hG^{a\pm}_{\mu\nu}G^{a\pm\mu\nu})$
\\ \hline

$h\gamma^\pm Z$
&1
&$[2{\bf3}][2{\bf3}]\,/\,\langle2{\bf3}\rangle\langle2{\bf3}\rangle$
&$h\gamma_{\mu\nu}^+Z^{+\mu\nu}\,/\,h\gamma_{\mu\nu}^-Z^{-\mu\nu}$
\\ \hline

$hZZ$\, ($hW^+W^-$)
&3
&$\bf \langle23\rangle[23],\,[23][23],\,\langle23\rangle\langle23\rangle$
&$hZ_{\mu}Z^{\mu},\,hZ_{\mu\nu}^+Z^{+\mu\nu},\,hZ_{\mu\nu}^-Z^{-\mu\nu}\,(\sim)$
\\ \hline \hline

$g^\pm g^\pm g^\pm$
&1
&$[12][23][31]\,/\,\langle12\rangle\langle23\rangle\langle31\rangle$
&$\epsilon_{abc} G_\mu^{a\pm\nu} G_\nu^{b\pm\rho} G_\rho^{c\pm\mu}$
\\ \hline

$\gamma^+ W^+W^-$
&2
&$[1 {\bf 2}][1{\bf3}]\langle {\bf 23}\rangle,\,[1{\bf 2}][1{\bf 3}][{\bf 23}]$
& $\gamma^{+\mu\nu} (W^+)_{\mu} (W^-)_{\nu},\,\gamma^{+\mu\nu} (W^+)_{\mu\rho}^+ (W^-)^{+\rho}_{\nu}$
\\ \hline
$\gamma^- W^+W^-$
&2
&$\vev{1 {\bf 2}} \vev{1{\bf3}}[ {\bf 23}],\,\vev{1{\bf 2}} \vev{1{\bf 3}} \vev{{\bf 23}}$
&$\gamma^{-\mu\nu} (W^+)_{\mu} (W^-)_{\nu},\,\gamma^{-\mu\nu} (W^+)_{\mu\rho}^- (W^-)^{-\rho}_{\nu}$
\\ \hline
$ZW^+W^-$ & 7 & $\{{\bf [12], \vev{12}}\} \times \{{\bf [23], \vev{23}}\}$ 
&$Z^{\pm}_{\mu\nu} (W^+)^{\pm\nu \rho} (W^-)^{\pm\mu\rho}$\\
&&$\times \{{\bf [31], \vev{31}}\}$ 
&$Z^{\pm}_ {\mu\nu} (W^+)^{\nu} (W^-)^{\mu},\,
Z_{\mu}(W^\pm)^{\rho} (W^\mp)^{\pm\mu}_\rho$ \\

&&\text{remove $\bf [12] [23] \vev{31}$}
& (remove the corresponding one)
\\ \hline

\end{tabular}
\caption{The $3$-pt bases of bosons. Notation $\sim$ means the bases are similar to the previous case.}
\end{table}

\section{Conclusion and Outlook} \label{sec:conclusion}
The use of the HEFT has proven valuable in describing the IR effects of new physics, such as the non-linear realization of electroweak symmetry. In HEFT, the unbroken gauge symmetry is limited to $U(1)_{\text{e.m.}} \times SU(3)_c$, which restricts all SM fields, except for photons and gluons, to be massive. Therefore, constructing HEFT operators regarding SM mass eigenstates is more natural and convenient for phenomenological studies. However, previous works have only constructed HEFT basis in terms of gauge eigenstates, resulting in basis that contain unphysical Goldstones. This type of HEFT is complicated for phenomenological studies, and its physical implications are not immediately clear.

In this work, we employed amplitude theory and the methods from~\cite{Dong:2021yak,Dong:2022mcv} to construct HEFT operators in Mathematica using SM mass eigenstates directly. We referred to this new formulation of HEFT as massive HEFT. In Section~\ref{sec:3pt} and Appendix~\ref{app:d8bases}, we presented the 3-point HEFT basis and 4-point basis up to dimension-8. Further research can explore the application of these bases to formulate the electroweak symmetry-breaking phase of the SMEFT and distinguish HEFT from SMEFT~\cite{Liu:2023jbq}.

Our method facilitates the construction of EFTs involving massive fields, which is generally a challenging task in traditional framework. Our method utilizes Mathematica codes and allows for the construction of generic types of EFTs, including dark matter EFTs and high-spin field EFTs, in a straightforward manner. Additionally, this method can enable the construction of the most comprehensive interactions between gravity and higher-spin particles, which is crucial in exploring the high-energy physics of gravity. The potential applications of this method in the construction of massive EFTs are vast and should be investigated in the future.

\section*{Acknowledgements}
This work is supported by the National Key Research and Development Program of
China under Grant No. 2021YFC2203004. T.M. is supported by ``Study in Israel" Fellowship for Outstanding Post-Doctoral Researchers from China and India by PBC of CHE and partially supported by grants from the NSF-BSF (No. 2018683), by the ISF (grant No. 482/20) and by the Azrieli foundation.  J.S. is
supported by the NSFC under Grants No. 12025507, No. 11690022, and No. 11947302, by
the Strategic Priority Research Program and Key Research Program of Frontier Science of the
Chinese Academy of Sciences (CAS) under Grants No. XDB21010200, No. XDB23010000,
No. XDPB15, and No. ZDBS-LY-7003, and by the CAS Project for Young Scientists in
Basic Research under Grant No. YSBR-006.

\clearpage
\appendix

\section{The four-point HEFT operators up to dim-8}
\label{app:d8bases}
In this appendix, we present the complete HEFT operators for some 4-point scattering processes involving SM particles up to dimension-8. In these bases, the Greek letters represent Lorentz indices while the English letters represent QCD color indices. The definitions of notation can be found around Eq.~(\ref{eq:matching}) and (\ref{eq:uugg_oprator}), with $D_\nu$ representing the covariant derivative.

Note that sometimes different operators correspond to the same amplitude bases, and we indicate them in parentheses. To obtain them explicitly, we merely need to substitute the fields in the EFT operators as suggested in parentheses.
The replacement rules are shown in the table below, where particles enclosed in the same braces can be interchanged with each other.
\begin{table}[H]
    \begin{center}
    \begin{tabular}{l|l}
    \hline
     \textbf{Lorentz structure}
     & \textbf{Fields} \\ \hline
     \text{left handed spinor} & $\{u_L,\,d_L,\,\bar{u}_R,\,\bar{d}_R\},\ \{e_L,\,\nu_L,\,\bar{e}_R,\,\bar{\nu}_R\}$ \\ \hline
     \text{right handed spinor} & $\{u_R,\,d_R,\,\bar{u}_L,\,\bar{d}_L\},\ \{e_R,\,\nu_R,\,\bar{e}_L,\,\bar{\nu}_L\}$\\ \hline
     \text{vector} & $\{W^+,\,W^-\}$\\ \hline
    \end{tabular}
    \end{center}
\end{table}

The following are some four-point scattering that are common on colliders and have important detection significance in particle physics:
\begin{itemize}
    \item[\ref{app:qqll}] e.g.~Drell-Yan process: The scattering process involving a quark-antiquark pair annihilating into a virtual photon or Z boson, which then decays into a lepton-antilepton pair, is an important channel for studying electroweak interactions and searching for new physics.

    \item[\ref{app:qqqq}] e.g.~Top quark pair production: The scattering process involving two top quarks being produced and subsequently decaying into multiple jets and leptons is a crucial probe of the top quark's properties and its role in electroweak symmetry breaking.

    \item[\ref{app:ZZhh}] Higgs boson production and decay: The scattering process involving two Higgs bosons decaying into two Z bosons is a significant channel for studying the properties of the Higgs boson at the LHC.
    
    \item[\ref{app:qqWW}] Top quark decay: The scattering process involving two top quarks decaying into two W bosons is a key channel for measuring the top quark mass and its coupling to the Higgs boson.

    \item[\ref{app:ZZZZ} \ref{app:ZZWW} \ref{app:WWWW}] Vector boson scattering: The scattering process involving two W or Z bosons scattering off each other is a pivotal test of the electroweak theory and has the potential to reveal new physics beyond the Standard Model.
\end{itemize}
The EFT operators of other scattering processes such as the annihilation of dark matter can be calculated using the program \url{https://github.com/zizhengzhou/MassiveAmplitude}, which also includes a PDF that lists EFT operators for all masses and spins up to dimension-8. In this context, we will only provide the operators for the abovementioned processes.

\input{result_new.tex}

\bibliographystyle{JHEP2}
\bibliography{AMP}

\end{document}

%% file: result_new.tex
\subsection{ $u \bar{u} \nu \bar{\nu}\,(d\bar{d}\nu\bar{\nu},\,u\bar{u}e^-e^+,\,d\bar{d}e^-e^+,\, d\bar{u}\nu e^+,\,u \bar{d}e^-\bar{\nu})$ }\label{app:qqll}\hspace*{\parindent}
$\bullet$ {Dimension = 6, $\mathcal{O}_{6}^{1\sim 10}$ }
\begin{longtable}[c]{|l|l|l|}
    \hline
    \multicolumn{3}{|c|}{ $u \bar{u} \nu \bar{\nu} \quad \pd=6 \quad \mathcal{O}_{6}^{1\sim 10}$} \\ \hline
    \endfirsthead %
    \endhead%
    \hline
    \endfoot%
    $ \displaystyle \left( {u}_{R}^{a} {\nu}_{R} \right) \left( {\bar{u}}_{L a} {\bar{\nu}}_{L} \right) $ & $ \displaystyle \left( {\bar{u}}_{L a} {\bar{\nu}}_{L} \right) \left( {u}_{L}^{a} {\nu}_{L} \right) $ & $ \displaystyle \left( {u}_{R}^{a} {\nu}_{R} \right) \left( {\bar{u}}_{R a} {\bar{\nu}}_{R} \right) $\\ \hline
    $ \displaystyle \left( {u}_{R}^{a} {\bar{u}}_{L a} \right) \left( {\nu}_{R} {\bar{\nu}}_{L} \right) $ & $ \displaystyle \left( {u}_{R}^{a} {\bar{\nu}}_{L} \right) \left( {\bar{u}}_{R a} {\nu}_{L} \right) $ & $ \displaystyle \left( {u}_{R}^{a} {\bar{u}}_{L a} \right) \left( {\nu}_{L} {\bar{\nu}}_{R} \right) $\\ \hline
    $ \displaystyle \left( {\nu}_{R} {\bar{\nu}}_{L} \right) \left( {u}_{L}^{a} {\bar{u}}_{R a} \right) $ & $ \displaystyle \left( {\bar{u}}_{L a} {\nu}_{R} \right) \left( {u}_{L}^{a} {\bar{\nu}}_{R} \right) $ & $ \displaystyle \left( {u}_{L}^{a} {\nu}_{L} \right) \left( {\bar{u}}_{R a} {\bar{\nu}}_{R} \right) $\\ \hline
    $ \displaystyle \left( {u}_{L}^{a} {\bar{u}}_{R a} \right) \left( {\nu}_{L} {\bar{\nu}}_{R} \right) $ &    &   \\ 
\end{longtable}
$\bullet$ {Dimension = 7, $\mathcal{O}_{7}^{1\sim 8}$ }
\begin{longtable}[c]{|l|l|l|}
    \hline
    \multicolumn{3}{|c|}{ $u \bar{u} \nu \bar{\nu} \quad \pd=7 \quad \mathcal{O}_{7}^{1\sim 8}$} \\ \hline
    \endfirsthead %
    \endhead%
    \hline
    \endfoot%
    $ \displaystyle \left( {\bar{u}}_{L a} \left( D_{\mu} {\bar{\nu}}_{L} \right) \right) \left( {\nu}_{R} \bar{\sigma}^{\mu} {u}_{L}^{a} \right) $ & $ \displaystyle \left( {u}_{R}^{a} \left( D_{\mu} {\bar{\nu}}_{L} \right) \right) \left( {\bar{u}}_{L a} \bar{\sigma}^{\mu} {\nu}_{L} \right) $ & $ \displaystyle \left( \left( D_{\mu} {\bar{u}}_{R a} \right) {\bar{\nu}}_{R} \right) \left( {u}_{R}^{a} \bar{\sigma}^{\mu} {\nu}_{L} \right) $\\ \hline
    $ \displaystyle \left( {u}_{R}^{a} \left( D_{\mu} {\bar{\nu}}_{L} \right) \right) \left( {\nu}_{R} \bar{\sigma}^{\mu} {\bar{u}}_{R a} \right) $ & $ \displaystyle \left( {u}_{R}^{a} \left( D_{\mu} {\nu}_{R} \right) \right) \left( {\bar{u}}_{L a} \bar{\sigma}^{\mu} {\bar{\nu}}_{R} \right) $ & $ \displaystyle \left( {\bar{u}}_{R a} \left( D_{\mu} {\nu}_{L} \right) \right) \left( {\bar{\nu}}_{L} \bar{\sigma}^{\mu} {u}_{L}^{a} \right) $\\ \hline
    $ \displaystyle \left( \left( D_{\mu} {\nu}_{L} \right) {\bar{\nu}}_{R} \right) \left( {\bar{u}}_{L a} \bar{\sigma}^{\mu} {u}_{L}^{a} \right) $ & $ \displaystyle \left( {\bar{u}}_{R a} \left( D_{\mu} {\bar{\nu}}_{R} \right) \right) \left( {\nu}_{R} \bar{\sigma}^{\mu} {u}_{L}^{a} \right) $ &   \\ 
\end{longtable}
$\bullet$ {Dimension = 8, $\mathcal{O}_{8}^{1\sim 18}$ }
\begin{longtable}[c]{|l|l|}
    \hline
    \multicolumn{2}{|c|}{ $u \bar{u} \nu \bar{\nu} \quad \pd=8 \quad \mathcal{O}_{8}^{1\sim 18}$} \\ \hline
    \endfirsthead %
    \endhead%
    \hline
    \endfoot%
    $ \displaystyle \left( {u}_{R}^{a} {\bar{\nu}}_{L} \right) \left( \left( D_{\nu} {\bar{u}}_{L a} \right) \bar{\sigma}^{\mu} \sigma^{\nu} \left( D_{\mu} {\nu}_{R} \right) \right) $ & $ \displaystyle \left( {u}_{L}^{a} {\nu}_{L} \right) \left( \left( D_{\nu} {\bar{u}}_{L a} \right) \bar{\sigma}^{\mu} \sigma^{\nu} \left( D_{\mu} {\bar{\nu}}_{L} \right) \right) $\\ \hline
    $ \displaystyle \left( {u}_{R}^{a} \left( D_{\mu} {\bar{\nu}}_{L} \right) \right) \left( \left( D_{\nu} {\bar{u}}_{L a} \right) \bar{\sigma}^{\mu} \sigma^{\nu} {\nu}_{R} \right) $ & $ \displaystyle \left( {\bar{u}}_{R a} \left( D_{\nu} {\nu}_{L} \right) \right) \left( {u}_{R}^{a} \bar{\sigma}^{\mu} \sigma^{\nu} \left( D_{\mu} {\bar{\nu}}_{L} \right) \right) $\\ \hline
    $ \displaystyle \left( {u}_{R}^{a} {\nu}_{R} \right) \left( \left( D_{\nu} {\bar{u}}_{L a} \right) \bar{\sigma}^{\mu} \sigma^{\nu} \left( D_{\mu} {\bar{\nu}}_{L} \right) \right) $ & $ \displaystyle \left( \left( D_{\nu} {\nu}_{L} \right) \left( D_{\mu} {\bar{\nu}}_{R} \right) \right) \left( {u}_{R}^{a} \bar{\sigma}^{\mu} \sigma^{\nu} {\bar{u}}_{L a} \right) $\\ \hline
    $ \displaystyle \left( \left( D_{\nu} {\bar{u}}_{L a} \right) \bar{\sigma}^{\mu} {u}_{L}^{a} \right) \left( {\bar{\nu}}_{L} \bar{\sigma}^{\nu} \left( D_{\mu} {\nu}_{L} \right) \right) $ & $ \displaystyle \left( {\bar{u}}_{L a} \left( D_{\nu} {\nu}_{R} \right) \right) \left( {u}_{L}^{a} \sigma^{\mu} \bar{\sigma}^{\nu} \left( D_{\mu} {\bar{\nu}}_{R} \right) \right) $\\ \hline
    $ \displaystyle \left( \left( D_{\nu} {\bar{u}}_{L a} \right) \bar{\sigma}^{\mu} {u}_{L}^{a} \right) \left( \left( D_{\mu} {\nu}_{R} \right) \bar{\sigma}^{\nu} {\bar{\nu}}_{R} \right) $ & $ \displaystyle \left( {u}_{R}^{a} {\nu}_{R} \right) \left( \left( D_{\nu} {\bar{u}}_{R a} \right) \sigma^{\mu} \bar{\sigma}^{\nu} \left( D_{\mu} {\bar{\nu}}_{R} \right) \right) $\\
\end{longtable}

\subsection{ $u \bar{u} d \bar{d}$ }\label{app:qqqq}\hspace*{\parindent}
$\bullet$ {Dimension = 6, $\mathcal{O}_{6}^{1\sim 20}$ }
\begin{longtable}[c]{|l|l|}
    \hline
    \multicolumn{2}{|c|}{ $u \bar{u} d \bar{d} \quad \pd=6 \quad \mathcal{O}_{6}^{1\sim 20}$} \\ \hline
    \endfirsthead %
    \endhead%
    \hline
    \endfoot%
    $ \displaystyle \left( {u}_{R}^{a} {d}_{R}^{b} \right) \left( {\bar{u}}_{L a} {\bar{d}}_{L b} \right) $ & $ \displaystyle \epsilon_{a c e} \epsilon_{b d f} \left( {u}_{R}^{a} {d}_{R}^{b} \right) \epsilon^{c d g} \left( {\bar{u}}_{L g} \epsilon^{e f h} {\bar{d}}_{L h} \right) $\\ \hline
    $ \displaystyle \left( {u}_{R}^{a} {\bar{u}}_{L a} \right) \left( {d}_{R}^{b} {\bar{d}}_{L b} \right) $ & $ \displaystyle \epsilon_{a c e} \epsilon_{b d f} \left( {u}_{R}^{a} \epsilon^{c d g} {\bar{u}}_{L g} \right) \left( {d}_{R}^{b} \epsilon^{e f h} {\bar{d}}_{L h} \right) $\\ \hline
    $ \displaystyle \left( {d}_{R}^{b} {\bar{d}}_{L b} \right) \left( {u}_{L}^{a} {\bar{u}}_{R a} \right) $ & $ \displaystyle \epsilon_{a c e} \epsilon_{b d f} \left( {d}_{R}^{b} \epsilon^{e f h} {\bar{d}}_{L h} \right) \left( {u}_{L}^{a} \epsilon^{c d g} {\bar{u}}_{R g} \right) $\\ \hline
    $ \displaystyle \left( {\bar{u}}_{L a} {\bar{d}}_{L b} \right) \left( {u}_{L}^{a} {d}_{L}^{b} \right) $ & $ \displaystyle \epsilon_{a c e} \epsilon_{b d f} \epsilon^{c d g} \left( {\bar{u}}_{L g} \epsilon^{e f h} {\bar{d}}_{L h} \right) \left( {u}_{L}^{a} {d}_{L}^{b} \right) $\\ \hline
    $ \displaystyle \left( {u}_{R}^{a} {\bar{d}}_{L b} \right) \left( {\bar{u}}_{R a} {d}_{L}^{b} \right) $ & $ \displaystyle \epsilon_{a c e} \epsilon_{b d f} \left( {u}_{R}^{a} \epsilon^{e f h} {\bar{d}}_{L h} \right) \epsilon^{c d g} \left( {\bar{u}}_{R g} {d}_{L}^{b} \right) $\\ \hline
    $ \displaystyle \left( {\bar{u}}_{L a} {d}_{R}^{b} \right) \left( {u}_{L}^{a} {\bar{d}}_{R b} \right) $ & $ \displaystyle \epsilon_{a c e} \epsilon_{b d f} \epsilon^{c d g} \left( {\bar{u}}_{L g} {d}_{R}^{b} \right) \left( {u}_{L}^{a} \epsilon^{e f h} {\bar{d}}_{R h} \right) $\\
\end{longtable}
$\bullet$ {Dimension = 7, $\mathcal{O}_{7}^{1\sim 16}$ }
\begin{longtable}[c]{|l|l|}
    \hline
    \multicolumn{2}{|c|}{ $u \bar{u} d \bar{d} \quad \pd=7 \quad \mathcal{O}_{7}^{1\sim 16}$} \\ \hline
    \endfirsthead %
    \endhead%
    \hline
    \endfoot%
    $ \displaystyle \left( {\bar{u}}_{L a} \left( D_{\mu} {\bar{d}}_{L b} \right) \right) \left( {d}_{R}^{b} \bar{\sigma}^{\mu} {u}_{L}^{a} \right) $ & $ \displaystyle \epsilon_{a c e} \epsilon_{b d f} \epsilon^{c d g} \left( {\bar{u}}_{L g} \left( D_{\mu} \epsilon^{e f h} {\bar{d}}_{L h} \right) \right) \left( {d}_{R}^{b} \bar{\sigma}^{\mu} {u}_{L}^{a} \right) $\\ \hline
    $ \displaystyle \left( {u}_{R}^{a} \left( D_{\mu} {\bar{d}}_{L b} \right) \right) \left( {d}_{R}^{b} \bar{\sigma}^{\mu} {\bar{u}}_{R a} \right) $ & $ \displaystyle \epsilon_{a c e} \epsilon_{b d f} \left( {u}_{R}^{a} \left( D_{\mu} \epsilon^{e f h} {\bar{d}}_{L h} \right) \right) \left( {d}_{R}^{b} \bar{\sigma}^{\mu} \epsilon^{c d g} {\bar{u}}_{R g} \right) $\\ \hline
    $ \displaystyle \left( {u}_{R}^{a} \left( D_{\mu} {\bar{d}}_{L b} \right) \right) \left( {\bar{u}}_{L a} \bar{\sigma}^{\mu} {d}_{L}^{b} \right) $ & $ \displaystyle \epsilon_{a c e} \epsilon_{b d f} \left( {u}_{R}^{a} \left( D_{\mu} \epsilon^{e f h} {\bar{d}}_{L h} \right) \right) \epsilon^{c d g} \left( {\bar{u}}_{L g} \bar{\sigma}^{\mu} {d}_{L}^{b} \right) $\\ \hline
    $ \displaystyle \left( {u}_{R}^{a} \left( D_{\mu} {d}_{R}^{b} \right) \right) \left( {\bar{u}}_{L a} \bar{\sigma}^{\mu} {\bar{d}}_{R b} \right) $ & $ \displaystyle \epsilon_{a c e} \epsilon_{b d f} \left( {u}_{R}^{a} \left( D_{\mu} {d}_{R}^{b} \right) \right) \epsilon^{c d g} \left( {\bar{u}}_{L g} \bar{\sigma}^{\mu} \epsilon^{e f h} {\bar{d}}_{R h} \right) $\\ \hline
    $ \displaystyle \left( \left( D_{\mu} {\bar{u}}_{R a} \right) {\bar{d}}_{R b} \right) \left( {u}_{R}^{a} \bar{\sigma}^{\mu} {d}_{L}^{b} \right) $ & $ \displaystyle \epsilon_{a c e} \epsilon_{b d f} \left( \left( D_{\mu} \epsilon^{c d g} {\bar{u}}_{R g} \right) \epsilon^{e f h} {\bar{d}}_{R h} \right) \left( {u}_{R}^{a} \bar{\sigma}^{\mu} {d}_{L}^{b} \right) $\\
\end{longtable}
$\bullet$ {Dimension = 8, $\mathcal{O}_{8}^{1\sim 36}$ }
\begin{longtable}[c]{|l|}
    \hline
    \multicolumn{1}{|c|}{ $u \bar{u} d \bar{d} \quad \pd=8 \quad \mathcal{O}_{8}^{1\sim 36}$} \\ \hline
    \endfirsthead %
    \endhead%
    \hline
    \endfoot%
    $ \displaystyle \left( {u}_{R}^{a} {\bar{d}}_{L b} \right) \left( \left( D_{\nu} {\bar{u}}_{L a} \right) \bar{\sigma}^{\mu} \sigma^{\nu} \left( D_{\mu} {d}_{R}^{b} \right) \right) $\\ \hline
    $ \displaystyle \left( {u}_{R}^{a} \left( D_{\mu} {\bar{d}}_{L b} \right) \right) \left( \left( D_{\nu} {\bar{u}}_{L a} \right) \bar{\sigma}^{\mu} \sigma^{\nu} {d}_{R}^{b} \right) $\\ \hline
    $ \displaystyle \left( {u}_{R}^{a} {d}_{R}^{b} \right) \left( \left( D_{\nu} {\bar{u}}_{L a} \right) \bar{\sigma}^{\mu} \sigma^{\nu} \left( D_{\mu} {\bar{d}}_{L b} \right) \right) $\\ \hline
    $ \displaystyle \left( \left( D_{\nu} {\bar{u}}_{L a} \right) \bar{\sigma}^{\mu} {u}_{L}^{a} \right) \left( {\bar{d}}_{L b} \bar{\sigma}^{\nu} \left( D_{\mu} {d}_{L}^{b} \right) \right) $\\ \hline
    $ \displaystyle \left( \left( D_{\nu} {\bar{u}}_{L a} \right) \bar{\sigma}^{\mu} {u}_{L}^{a} \right) \left( \left( D_{\mu} {d}_{R}^{b} \right) \bar{\sigma}^{\nu} {\bar{d}}_{R b} \right) $\\ \hline
    $ \displaystyle \left( \left( D_{\nu} {\bar{u}}_{R a} \right) \left( D_{\mu} {d}_{L}^{b} \right) \right) \left( {u}_{R}^{a} \bar{\sigma}^{\mu} \sigma^{\nu} {\bar{d}}_{L b} \right) $\\ \hline
    $ \displaystyle \left( \left( D_{\nu} {\bar{u}}_{R a} \right) {\bar{d}}_{R b} \right) \left( {u}_{R}^{a} \bar{\sigma}^{\mu} \sigma^{\nu} \left( D_{\mu} {d}_{R}^{b} \right) \right) $\\ \hline
    $ \displaystyle \left( \left( D_{\nu} {d}_{R}^{b} \right) \bar{\sigma}^{\mu} {\bar{u}}_{R a} \right) \left( \left( D_{\mu} {\bar{d}}_{L b} \right) \bar{\sigma}^{\nu} {u}_{L}^{a} \right) $\\
\end{longtable}

\subsection{ $Z Z h h$ }\label{app:ZZhh}\hspace*{\parindent}
$\bullet$ {Dimension = 4, $\mathcal{O}_{4}^{1}$ }
\begin{longtable}[c]{|l|}
    \hline
    \multicolumn{1}{|c|}{ $Z Z h h \quad \pd=4 \quad \mathcal{O}_{4}^{1}$} \\ \hline
    \endfirsthead %
    \endhead%
    \hline
    \endfoot%
    $ \displaystyle {Z}_{ \mu} {Z}_{ \nu} h h \operatorname{Tr} \left( \sigma^{\mu} \bar{\sigma}^{\nu} \right) $\\ 
\end{longtable}
$\bullet$ {Dimension = 6, $\mathcal{O}_{6}^{1\sim 4}$ }
\begin{longtable}[c]{|l|l|}
    \hline
    \multicolumn{2}{|c|}{ $Z Z h h \quad \pd=6 \quad \mathcal{O}_{6}^{1\sim 4}$} \\ \hline
    \endfirsthead %
    \endhead%
    \hline
    \endfoot%
    $ \displaystyle {Z}^{+}_{ \mu \nu} {Z}^{+}_{ \rho \sigma} h h \operatorname{Tr} \left( \bar{\sigma}^{\mu \nu} \bar{\sigma}^{\rho \sigma} \right) $ & $ \displaystyle {Z}_{ \rho} {Z}_{ \sigma} \left( D_{\nu} h \right) \left( D_{\mu} h \right) \operatorname{Tr} \left( \bar{\sigma}^{\mu} \sigma^{\nu} \bar{\sigma}^{\sigma} \sigma^{\rho} \right) $\\ \hline
    $ \displaystyle {Z}^{-}_{ \mu \nu} {Z}^{-}_{ \rho \sigma} h h \operatorname{Tr} \left( \sigma^{\mu \nu} \sigma^{\rho \sigma} \right) $ & $ \displaystyle {Z}_{ \nu} {Z}_{ \sigma} \left( D_{\mu} D_{\rho} h \right) h \operatorname{Tr} \left( \bar{\sigma}^{\mu} \sigma^{\nu} \right) \operatorname{Tr} \left( \bar{\sigma}^{\rho} \sigma^{\sigma} \right) $\\ 
\end{longtable}
$\bullet$ {Dimension = 8, $\mathcal{O}_{8}^{1\sim 7}$ }
\begin{longtable}[c]{|l|}
    \hline
    \multicolumn{1}{|c|}{ $Z Z h h \quad \pd=8 \quad \mathcal{O}_{8}^{1\sim 7}$} \\ \hline
    \endfirsthead %
    \endhead%
    \hline
    \endfoot%
    $ \displaystyle {Z}^{-}_{ \rho \sigma} {Z}^{+}_{ \xi \tau} \left( D_{\mu} D_{\nu} h \right) h \operatorname{Tr} \left( \bar{\sigma}^{\mu} \sigma^{\rho \sigma} \bar{\sigma}^{\nu} \bar{\sigma}^{\xi \tau} \right) $\\ \hline
    $ \displaystyle {Z}^{+}_{ \rho \sigma} {Z}^{+}_{ \xi \tau} \left( D_{\nu} h \right) \left( D_{\mu} h \right) \operatorname{Tr} \left( \bar{\sigma}^{\mu} \sigma^{\nu} \bar{\sigma}^{\xi \tau} \bar{\sigma}^{\rho \sigma} \right) $\\ \hline
    $ \displaystyle {Z}^{-}_{ \rho \sigma} {Z}^{-}_{ \xi \tau} \left( D_{\mu} h \right) \left( D_{\nu} h \right) \operatorname{Tr} \left( \sigma^{\rho \sigma} \sigma^{\mu} \bar{\sigma}^{\nu} \sigma^{\xi \tau} \right) $\\ \hline
    $ \displaystyle {Z}_{ \sigma} {Z}^{+}_{ \xi \tau} \left( D_{\nu} D_{\rho} h \right) \left( D_{\mu} h \right) \operatorname{Tr} \left( \bar{\sigma}^{\mu} \sigma^{\nu} \bar{\sigma}^{\xi \tau} \bar{\sigma}^{\rho} \sigma^{\sigma} \right) $\\ \hline
    $ \displaystyle {Z}^{-}_{ \sigma \xi} {Z}_{ \tau} \left( D_{\mu} D_{\nu} h \right) \left( D_{\rho} h \right) \operatorname{Tr} \left( \bar{\sigma}^{\mu} \sigma^{\sigma \xi} \sigma^{\nu} \bar{\sigma}^{\rho} \sigma^{\tau} \right) $\\ \hline
    $ \displaystyle {Z}_{ \xi} {Z}_{ \tau} \left( D_{\nu} D_{\sigma} h \right) \left( D_{\mu} D_{\rho} h \right) \operatorname{Tr} \left( \bar{\sigma}^{\mu} \sigma^{\nu} \bar{\sigma}^{\rho} \sigma^{\sigma} \bar{\sigma}^{\tau} \sigma^{\xi} \right) $\\ \hline
    $ \displaystyle {Z}_{ \nu} \left( D_{\sigma} {Z}_{ \tau} \right) \left( D_{\mu} D_{\rho} h \right) \left( D_{\xi} h \right) \operatorname{Tr} \left( \bar{\sigma}^{\mu} \sigma^{\nu} \right) \operatorname{Tr} \left( \bar{\sigma}^{\rho} \sigma^{\sigma} \bar{\sigma}^{\xi} \sigma^{\tau} \right) $\\ 
\end{longtable}

\subsection{ $u \bar{u} W^+ W^-\,(d \bar{d} W^+ W^-)$ }\label{app:qqWW}\hspace*{\parindent}
$\bullet$ {Dimension = 5, $\mathcal{O}_{5}^{1\sim 4}$ }
\begin{longtable}[c]{|l|l|l|}
    \hline
    \multicolumn{3}{|c|}{ $u \bar{u} W^+ W^- \quad \pd=5 \quad \mathcal{O}_{5}^{1\sim 4}$} \\ \hline
    \endfirsthead %
    \endhead%
    \hline
    \endfoot%
    $ \displaystyle {{(W^+)}}_{ \mu} {{(W^-)}}_{ \nu} \left( {u}_{R}^{a} \bar{\sigma}^{\mu} \bar{\sigma}^{\nu} {\bar{u}}_{L a} \right) $ & $ \displaystyle {{(W^+)}}_{ \mu} {{(W^-)}}_{ \nu} \left( {u}_{L}^{a} \sigma^{\mu} \bar{\sigma}^{\nu} {\bar{u}}_{R a} \right) $ & $ \displaystyle {{(W^+)}}_{ \mu} {{(W^-)}}_{ \nu} \left( {u}_{R}^{a} {\bar{u}}_{L a} \right) \operatorname{Tr} \left( \sigma^{\mu} \bar{\sigma}^{\nu} \right) $\\ \hline
    $ \displaystyle {{(W^+)}}_{ \mu} {{(W^-)}}_{ \nu} \left( {u}_{L}^{a} {\bar{u}}_{R a} \right) \operatorname{Tr} \left( \sigma^{\mu} \bar{\sigma}^{\nu} \right) $ &    &   \\ 
\end{longtable}
$\bullet$ {Dimension = 6, $\mathcal{O}_{6}^{1\sim 12}$ }
\begin{longtable}[c]{|l|l|}
    \hline
    \multicolumn{2}{|c|}{ $u \bar{u} W^+ W^- \quad \pd=6 \quad \mathcal{O}_{6}^{1\sim 12}$} \\ \hline
    \endfirsthead %
    \endhead%
    \hline
    \endfoot%
    $ \displaystyle {{(W^+)}}_{ \mu} {{(W^-)}}^{+}_{ \nu \rho} \left( {\bar{u}}_{L a} \bar{\sigma}^{\nu \rho} \sigma^{\mu} {u}_{L}^{a} \right) $ & $ \displaystyle {{(W^+)}}_{ \mu} {{(W^-)}}^{-}_{ \nu \rho} \left( {\bar{u}}_{L a} \bar{\sigma}^{\mu} \sigma^{\nu \rho} {u}_{L}^{a} \right) $\\ \hline
    $ \displaystyle {{(W^+)}}_{ \mu} {{(W^-)}}^{+}_{ \nu \rho} \left( {u}_{R}^{a} \bar{\sigma}^{\nu \rho} \sigma^{\mu} {\bar{u}}_{R a} \right) $ & $ \displaystyle {{(W^+)}}_{ \mu} {{(W^-)}}^{-}_{ \nu \rho} \left( {u}_{R}^{a} \bar{\sigma}^{\mu} \sigma^{\nu \rho} {\bar{u}}_{R a} \right) $\\ \hline
    $ \displaystyle {{(W^+)}}^{+}_{ \mu \nu} {{(W^-)}}_{ \rho} \left( {\bar{u}}_{L a} \bar{\sigma}^{\mu \nu} \bar{\sigma}^{\rho} {u}_{L}^{a} \right) $ & $ \displaystyle {{(W^+)}}_{ \nu} {{(W^-)}}_{ \rho} \left( \left( D_{\mu} {\bar{u}}_{L a} \right) \bar{\sigma}^{\rho} \bar{\sigma}^{\mu} \sigma^{\nu} {u}_{L}^{a} \right) $\\ \hline
    $ \displaystyle {{(W^+)}}^{+}_{ \mu \nu} {{(W^-)}}_{ \rho} \left( {u}_{R}^{a} \bar{\sigma}^{\mu \nu} \bar{\sigma}^{\rho} {\bar{u}}_{R a} \right) $ & $ \displaystyle {{(W^+)}}_{ \nu} \left( D_{\mu} {{(W^-)}}_{ \rho} \right) \left( {\bar{u}}_{L a} \bar{\sigma}^{\rho} \sigma^{\nu} \bar{\sigma}^{\mu} {u}_{L}^{a} \right) $\\ \hline
    $ \displaystyle {{(W^+)}}^{-}_{ \mu \nu} {{(W^-)}}_{ \rho} \left( {\bar{u}}_{L a} \bar{\sigma}^{\rho} \sigma^{\mu \nu} {u}_{L}^{a} \right) $ & $ \displaystyle {{(W^+)}}_{ \nu} \left( D_{\mu} {{(W^-)}}_{ \rho} \right) \left( {u}_{R}^{a} \bar{\sigma}^{\rho} \sigma^{\nu} \bar{\sigma}^{\mu} {\bar{u}}_{R a} \right) $\\ \hline
    $ \displaystyle {{(W^+)}}^{-}_{ \mu \nu} {{(W^-)}}_{ \rho} \left( {u}_{R}^{a} \bar{\sigma}^{\rho} \sigma^{\mu \nu} {\bar{u}}_{R a} \right) $ & $ \displaystyle {{(W^+)}}_{ \nu} {{(W^-)}}_{ \rho} \left( {u}_{R}^{a} \bar{\sigma}^{\rho} \left( D_{\mu} {\bar{u}}_{R a} \right) \right) \operatorname{Tr} \left( \bar{\sigma}^{\mu} \sigma^{\nu} \right) $\\
\end{longtable}
$\bullet$ {Dimension = 7, $\mathcal{O}_{7}^{1\sim 24}$ }
\begin{longtable}[c]{|l|l|}
    \hline
    \multicolumn{2}{|c|}{ $u \bar{u} W^+ W^- \quad \pd=7 \quad \mathcal{O}_{7}^{1\sim 24}$} \\ \hline
    \endfirsthead %
    \endhead%
    \hline
    \endfoot%
    $ \displaystyle {{(W^+)}}^{+}_{ \mu \nu} {{(W^-)}}^{+}_{ \rho \sigma} \left( {u}_{R}^{a} \bar{\sigma}^{\mu \nu} \bar{\sigma}^{\rho \sigma} {\bar{u}}_{L a} \right) $ & $ \displaystyle {{(W^+)}}_{ \nu} {{(W^-)}}^{-}_{ \rho \sigma} \left( {u}_{L}^{a} \sigma^{\nu} \bar{\sigma}^{\mu} \sigma^{\rho \sigma} \left( D_{\mu} {\bar{u}}_{R a} \right) \right) $\\ \hline
    $ \displaystyle {{(W^+)}}^{-}_{ \mu \nu} {{(W^-)}}^{-}_{ \rho \sigma} \left( {u}_{L}^{a} \sigma^{\mu \nu} \sigma^{\rho \sigma} {\bar{u}}_{R a} \right) $ & $ \displaystyle \left( D_{\mu} {{(W^+)}}^{-}_{ \nu \rho} \right) {{(W^-)}}_{ \sigma} \left( {u}_{L}^{a} \bar{\sigma}^{\mu} \sigma^{\sigma} \sigma^{\nu \rho} {\bar{u}}_{R a} \right) $\\ \hline
    $ \displaystyle {{(W^+)}}^{+}_{ \mu \nu} {{(W^-)}}^{+}_{ \rho \sigma} \left( {u}_{R}^{a} {\bar{u}}_{L a} \right) \operatorname{Tr} \left( \bar{\sigma}^{\mu \nu} \bar{\sigma}^{\rho \sigma} \right) $ & $ \displaystyle {{(W^+)}}_{ \nu} \left( D_{\mu} {{(W^-)}}^{-}_{ \rho \sigma} \right) \left( {u}_{L}^{a} \bar{\sigma}^{\mu} \sigma^{\nu} \sigma^{\rho \sigma} {\bar{u}}_{R a} \right) $\\ \hline
    $ \displaystyle {{(W^+)}}^{+}_{ \mu \nu} {{(W^-)}}^{+}_{ \rho \sigma} \left( {u}_{L}^{a} {\bar{u}}_{R a} \right) \operatorname{Tr} \left( \bar{\sigma}^{\mu \nu} \bar{\sigma}^{\rho \sigma} \right) $ & $ \displaystyle {{(W^+)}}_{ \nu} {{(W^-)}}^{+}_{ \rho \sigma} \left( {u}_{R}^{a} \bar{\sigma}^{\rho \sigma} \left( D_{\mu} {\bar{u}}_{L a} \right) \right) \operatorname{Tr} \left( \bar{\sigma}^{\mu} \sigma^{\nu} \right) $\\ \hline
    $ \displaystyle {{(W^+)}}^{+}_{ \nu \rho} {{(W^-)}}_{ \sigma} \left( {u}_{R}^{a} \bar{\sigma}^{\nu \rho} \bar{\sigma}^{\mu} \bar{\sigma}^{\sigma} \left( D_{\mu} {\bar{u}}_{L a} \right) \right) $ & $ \displaystyle {{(W^+)}}_{ \nu} \left( D_{\mu} {{(W^-)}}^{+}_{ \rho \sigma} \right) \left( {u}_{R}^{a} \bar{\sigma}^{\rho \sigma} {\bar{u}}_{L a} \right) \operatorname{Tr} \left( \bar{\sigma}^{\mu} \sigma^{\nu} \right) $\\
\end{longtable}
$\bullet$ {Dimension = 8, $\mathcal{O}_{8}^{1\sim 30}$ }
\begin{longtable}[c]{|l|}
    \hline
    \multicolumn{1}{|c|}{ $u \bar{u} W^+ W^- \quad \pd=8 \quad \mathcal{O}_{8}^{1\sim 30}$} \\ \hline
    \endfirsthead %
    \endhead%
    \hline
    \endfoot%
    $ \displaystyle {{(W^+)}}^{+}_{ \nu \rho} \left( D_{\mu} {{(W^-)}}^{+}_{ \sigma \xi} \right) \left( {\bar{u}}_{L a} \bar{\sigma}^{\sigma \xi} \bar{\sigma}^{\nu \rho} \bar{\sigma}^{\mu} {u}_{L}^{a} \right) $\\ \hline
    $ \displaystyle {{(W^+)}}^{+}_{ \nu \rho} \left( D_{\mu} {{(W^-)}}^{+}_{ \sigma \xi} \right) \left( {u}_{R}^{a} \bar{\sigma}^{\sigma \xi} \bar{\sigma}^{\nu \rho} \bar{\sigma}^{\mu} {\bar{u}}_{R a} \right) $\\ \hline
    $ \displaystyle {{(W^+)}}^{-}_{ \nu \rho} {{(W^-)}}^{+}_{ \sigma \xi} \left( \left( D_{\mu} {\bar{u}}_{L a} \right) \bar{\sigma}^{\sigma \xi} \bar{\sigma}^{\mu} \sigma^{\nu \rho} {u}_{L}^{a} \right) $\\ \hline
    $ \displaystyle {{(W^+)}}^{-}_{ \nu \rho} {{(W^-)}}^{+}_{ \sigma \xi} \left( {u}_{R}^{a} \bar{\sigma}^{\sigma \xi} \bar{\sigma}^{\mu} \sigma^{\nu \rho} \left( D_{\mu} {\bar{u}}_{R a} \right) \right) $\\ \hline
    $ \displaystyle {{(W^+)}}^{+}_{ \nu \rho} {{(W^-)}}^{-}_{ \sigma \xi} \left( \left( D_{\mu} {\bar{u}}_{L a} \right) \bar{\sigma}^{\nu \rho} \bar{\sigma}^{\mu} \sigma^{\sigma \xi} {u}_{L}^{a} \right) $\\ \hline
    $ \displaystyle {{(W^+)}}^{+}_{ \nu \rho} {{(W^-)}}^{-}_{ \sigma \xi} \left( {u}_{R}^{a} \bar{\sigma}^{\nu \rho} \bar{\sigma}^{\mu} \sigma^{\sigma \xi} \left( D_{\mu} {\bar{u}}_{R a} \right) \right) $\\ \hline
    $ \displaystyle {{(W^+)}}^{-}_{ \nu \rho} {{(W^-)}}^{-}_{ \sigma \xi} \left( {u}_{R}^{a} \bar{\sigma}^{\mu} \sigma^{\nu \rho} \sigma^{\sigma \xi} \left( D_{\mu} {\bar{u}}_{R a} \right) \right) $\\ \hline
    $ \displaystyle {{(W^+)}}^{-}_{ \rho \sigma} {{(W^-)}}_{ \xi} \left( {u}_{R}^{a} \bar{\sigma}^{\mu} \sigma^{\rho \sigma} \bar{\sigma}^{\nu} \sigma^{\xi} \left( D_{\mu} D_{\nu} {\bar{u}}_{R a} \right) \right) $\\
\end{longtable}

\subsection{ $Z Z Z Z$ }\label{app:ZZZZ}\hspace*{\parindent}
$\bullet$ {Dimension = 4, $\mathcal{O}_{4}^{1}$ }
\begin{longtable}[c]{|l|}
    \hline
    \multicolumn{1}{|c|}{ $Z Z Z Z \quad \pd=4 \quad \mathcal{O}_{4}^{1}$} \\ \hline
    \endfirsthead %
    \endhead%
    \hline
    \endfoot%
    $ \displaystyle {Z}_{ \mu} {Z}_{ \nu} {Z}_{ \rho} {Z}_{ \sigma} \operatorname{Tr} \left( \sigma^{\mu} \bar{\sigma}^{\rho} \right) \operatorname{Tr} \left( \sigma^{\nu} \bar{\sigma}^{\sigma} \right) $\\ 
\end{longtable}
$\bullet$ {Dimension = 6, $\mathcal{O}_{6}^{1\sim 4}$ }
\begin{longtable}[c]{|l|}
    \hline
    \multicolumn{1}{|c|}{ $Z Z Z Z \quad \pd=6 \quad \mathcal{O}_{6}^{1\sim 4}$} \\ \hline
    \endfirsthead %
    \endhead%
    \hline
    \endfoot%
    $ \displaystyle {Z}_{ \mu} {Z}_{ \nu} {Z}^{+}_{ \rho \sigma} {Z}^{+}_{ \xi \tau} \operatorname{Tr} \left( \sigma^{\mu} \sigma^{\nu} \bar{\sigma}^{\xi \tau} \bar{\sigma}^{\rho \sigma} \right) $\\ \hline
    $ \displaystyle {Z}^{-}_{ \mu \nu} {Z}_{ \rho} {Z}_{ \sigma} {Z}^{+}_{ \xi \tau} \operatorname{Tr} \left( \sigma^{\rho} \sigma^{\mu \nu} \sigma^{\sigma} \bar{\sigma}^{\xi \tau} \right) $\\ \hline
    $ \displaystyle {Z}^{-}_{ \mu \nu} {Z}^{-}_{ \rho \sigma} {Z}_{ \xi} {Z}_{ \tau} \operatorname{Tr} \left( \sigma^{\xi} \sigma^{\mu \nu} \sigma^{\rho \sigma} \bar{\sigma}^{\tau} \right) $\\ \hline
    $ \displaystyle \left( D_{\mu} {Z}_{ \rho} \right) {Z}_{ \sigma} \left( D_{\nu} {Z}_{ \xi} \right) {Z}_{ \tau} \operatorname{Tr} \left( \sigma^{\sigma} \bar{\sigma}^{\mu} \sigma^{\xi} \bar{\sigma}^{\tau} \right) \operatorname{Tr} \left( \bar{\sigma}^{\nu} \sigma^{\rho} \right) $\\ 
\end{longtable}
$\bullet$ {Dimension = 8, $\mathcal{O}_{8}^{1\sim 9}$ }
\begin{longtable}[c]{|l|}
    \hline
    \multicolumn{1}{|c|}{ $Z Z Z Z \quad \pd=8 \quad \mathcal{O}_{8}^{1\sim 9}$} \\ \hline
    \endfirsthead %
    \endhead%
    \hline
    \endfoot%
    $ \displaystyle {Z}^{+}_{ \mu \nu} {Z}^{+}_{ \rho \sigma} {Z}^{+}_{ \xi \tau} {Z}^{+}_{ \zeta \eta} \operatorname{Tr} \left( \bar{\sigma}^{\mu \nu} \bar{\sigma}^{\xi \tau} \right) \operatorname{Tr} \left( \bar{\sigma}^{\rho \sigma} \bar{\sigma}^{\zeta \eta} \right) $\\ \hline
    $ \displaystyle {Z}^{-}_{ \mu \nu} {Z}^{-}_{ \rho \sigma} {Z}^{+}_{ \xi \tau} {Z}^{+}_{ \zeta \eta} \operatorname{Tr} \left( \bar{\sigma}^{\xi \tau} \bar{\sigma}^{\zeta \eta} \right) \operatorname{Tr} \left( \sigma^{\mu \nu} \sigma^{\rho \sigma} \right) $\\ \hline
    $ \displaystyle {Z}^{-}_{ \mu \nu} {Z}^{-}_{ \rho \sigma} {Z}^{-}_{ \xi \tau} {Z}^{-}_{ \zeta \eta} \operatorname{Tr} \left( \sigma^{\mu \nu} \sigma^{\xi \tau} \right) \operatorname{Tr} \left( \sigma^{\rho \sigma} \sigma^{\zeta \eta} \right) $\\ \hline
    $ \displaystyle {Z}_{ \rho} {Z}_{ \sigma} {Z}^{+}_{ \xi \tau} \left( D_{\mu} D_{\nu} {Z}^{+}_{ \zeta \eta} \right) \operatorname{Tr} \left( \sigma^{\rho} \bar{\sigma}^{\mu} \bar{\sigma}^{\xi \tau} \bar{\sigma}^{\nu} \sigma^{\sigma} \bar{\sigma}^{\zeta \eta} \right) $\\ \hline
    $ \displaystyle {Z}^{-}_{ \rho \sigma} \left( D_{\nu} {Z}^{+}_{ \xi \tau} \right) \left( D_{\mu} {Z}_{ \zeta} \right) {Z}_{ \eta} \operatorname{Tr} \left( \bar{\sigma}^{\mu} \sigma^{\rho \sigma} \sigma^{\zeta} \bar{\sigma}^{\nu} \bar{\sigma}^{\eta} \bar{\sigma}^{\xi \tau} \right) $\\ \hline
    $ \displaystyle {Z}^{-}_{ \rho \sigma} \left( D_{\nu} {Z}_{ \xi} \right) \left( D_{\mu} {Z}^{-}_{ \tau \zeta} \right) {Z}_{ \eta} \operatorname{Tr} \left( \bar{\sigma}^{\mu} \sigma^{\rho \sigma} \sigma^{\tau \zeta} \bar{\sigma}^{\nu} \sigma^{\eta} \sigma^{\xi} \right) $\\ \hline
    $ \displaystyle {Z}^{-}_{ \rho \sigma} \left( D_{\nu} {Z}_{ \xi} \right) \left( D_{\mu} {Z}_{ \tau} \right) {Z}^{-}_{ \zeta \eta} \operatorname{Tr} \left( \bar{\sigma}^{\mu} \sigma^{\rho \sigma} \sigma^{\tau} \bar{\sigma}^{\nu} \sigma^{\zeta \eta} \sigma^{\xi} \right) $\\
\end{longtable}

\subsection{ $Z Z W^+ W^-$ }\label{app:ZZWW}\hspace*{\parindent}
$\bullet$ {Dimension = 4, $\mathcal{O}_{4}^{1\sim 2}$ }
\begin{longtable}[c]{|l|l|}
    \hline
    \multicolumn{2}{|c|}{ $Z Z W^+ W^- \quad \pd=4 \quad \mathcal{O}_{4}^{1\sim 2}$} \\ \hline
    \endfirsthead %
    \endhead%
    \hline
    \endfoot%
    $ \displaystyle {Z}_{ \mu} {Z}_{ \nu} {{(W^+)}}_{ \rho} {{(W^-)}}_{ \sigma} \operatorname{Tr} \left( \sigma^{\mu} \sigma^{\nu} \bar{\sigma}^{\sigma} \bar{\sigma}^{\rho} \right) $ & $ \displaystyle {Z}_{ \mu} {Z}_{ \nu} {{(W^+)}}_{ \rho} {{(W^-)}}_{ \sigma} \operatorname{Tr} \left( \sigma^{\mu} \bar{\sigma}^{\rho} \right) \operatorname{Tr} \left( \sigma^{\nu} \bar{\sigma}^{\sigma} \right) $\\ 
\end{longtable}
$\bullet$ {Dimension = 6, $\mathcal{O}_{6}^{1\sim 29}$ }
\begin{longtable}[c]{|l|}
    \hline
    \multicolumn{1}{|c|}{ $Z Z W^+ W^- \quad \pd=6 \quad \mathcal{O}_{6}^{1\sim 29}$} \\ \hline
    \endfirsthead %
    \endhead%
    \hline
    \endfoot%
    $ \displaystyle {Z}_{ \mu} {Z}_{ \nu} {{(W^+)}}^{+}_{ \rho \sigma} {{(W^-)}}^{+}_{ \xi \tau} \operatorname{Tr} \left( \sigma^{\mu} \sigma^{\nu} \bar{\sigma}^{\xi \tau} \bar{\sigma}^{\rho \sigma} \right) $\\ \hline
    $ \displaystyle {Z}^{+}_{ \mu \nu} {Z}_{ \rho} {{(W^+)}}_{ \sigma} {{(W^-)}}^{+}_{ \xi \tau} \operatorname{Tr} \left( \bar{\sigma}^{\mu \nu} \bar{\sigma}^{\sigma} \sigma^{\rho} \bar{\sigma}^{\xi \tau} \right) $\\ \hline
    $ \displaystyle {Z}_{ \mu} {Z}^{+}_{ \nu \rho} {{(W^+)}}^{+}_{ \sigma \xi} {{(W^-)}}_{ \tau} \operatorname{Tr} \left( \sigma^{\mu} \bar{\sigma}^{\tau} \bar{\sigma}^{\nu \rho} \bar{\sigma}^{\sigma \xi} \right) $\\ \hline
    $ \displaystyle {Z}^{+}_{ \mu \nu} {Z}^{+}_{ \rho \sigma} {{(W^+)}}_{ \xi} {{(W^-)}}_{ \tau} \operatorname{Tr} \left( \bar{\sigma}^{\mu \nu} \bar{\sigma}^{\rho \sigma} \bar{\sigma}^{\tau} \bar{\sigma}^{\xi} \right) $\\ \hline
    $ \displaystyle {Z}^{-}_{ \mu \nu} {Z}_{ \rho} {{(W^+)}}_{ \sigma} {{(W^-)}}^{+}_{ \xi \tau} \operatorname{Tr} \left( \sigma^{\rho} \sigma^{\mu \nu} \sigma^{\sigma} \bar{\sigma}^{\xi \tau} \right) $\\ \hline
    $ \displaystyle {Z}_{ \mu} {Z}_{ \nu} {{(W^+)}}^{-}_{ \rho \sigma} {{(W^-)}}^{+}_{ \xi \tau} \operatorname{Tr} \left( \sigma^{\mu} \sigma^{\rho \sigma} \sigma^{\nu} \bar{\sigma}^{\xi \tau} \right) $\\ \hline
    $ \displaystyle {Z}^{-}_{ \mu \nu} {Z}_{ \rho} {{(W^+)}}^{+}_{ \sigma \xi} {{(W^-)}}_{ \tau} \operatorname{Tr} \left( \sigma^{\rho} \sigma^{\mu \nu} \bar{\sigma}^{\tau} \bar{\sigma}^{\sigma \xi} \right) $\\ \hline
    $ \displaystyle {Z}^{-}_{ \mu \nu} {Z}^{+}_{ \rho \sigma} {{(W^+)}}_{ \xi} {{(W^-)}}_{ \tau} \operatorname{Tr} \left( \bar{\sigma}^{\rho \sigma} \bar{\sigma}^{\xi} \sigma^{\mu \nu} \bar{\sigma}^{\tau} \right) $\\
\end{longtable}
$\bullet$ {Dimension = 8, $\mathcal{O}_{8}^{1\sim 71}$ }
\begin{longtable}[c]{|l|}
    \hline
    \multicolumn{1}{|c|}{ $Z Z W^+ W^- \quad \pd=8 \quad \mathcal{O}_{8}^{1\sim 71}$} \\ \hline
    \endfirsthead %
    \endhead%
    \hline
    \endfoot%
    $ \displaystyle {Z}^{+}_{ \mu \nu} {Z}^{+}_{ \rho \sigma} {{(W^+)}}^{+}_{ \xi \tau} {{(W^-)}}^{+}_{ \zeta \eta} \operatorname{Tr} \left( \bar{\sigma}^{\mu \nu} \bar{\sigma}^{\rho \sigma} \bar{\sigma}^{\zeta \eta} \bar{\sigma}^{\xi \tau} \right) $\\ \hline
    $ \displaystyle {Z}^{-}_{ \mu \nu} {Z}^{-}_{ \rho \sigma} {{(W^+)}}^{-}_{ \xi \tau} {{(W^-)}}^{-}_{ \zeta \eta} \operatorname{Tr} \left( \sigma^{\mu \nu} \sigma^{\xi \tau} \sigma^{\zeta \eta} \sigma^{\rho \sigma} \right) $\\ \hline
    $ \displaystyle {Z}^{+}_{ \mu \nu} {Z}^{+}_{ \rho \sigma} {{(W^+)}}^{+}_{ \xi \tau} {{(W^-)}}^{+}_{ \zeta \eta} \operatorname{Tr} \left( \bar{\sigma}^{\mu \nu} \bar{\sigma}^{\xi \tau} \right) \operatorname{Tr} \left( \bar{\sigma}^{\rho \sigma} \bar{\sigma}^{\zeta \eta} \right) $\\ \hline
    $ \displaystyle {Z}^{+}_{ \nu \rho} {Z}_{ \sigma} {{(W^+)}}^{+}_{ \xi \tau} \left( D_{\mu} {{(W^-)}}^{+}_{ \zeta \eta} \right) \operatorname{Tr} \left( \bar{\sigma}^{\nu \rho} \bar{\sigma}^{\xi \tau} \bar{\sigma}^{\mu} \sigma^{\sigma} \bar{\sigma}^{\zeta \eta} \right) $\\ \hline
    $ \displaystyle {Z}^{+}_{ \nu \rho} {Z}^{+}_{ \sigma \xi} {{(W^+)}}_{ \tau} \left( D_{\mu} {{(W^-)}}^{+}_{ \zeta \eta} \right) \operatorname{Tr} \left( \bar{\sigma}^{\nu \rho} \bar{\sigma}^{\tau} \bar{\sigma}^{\mu} \bar{\sigma}^{\sigma \xi} \bar{\sigma}^{\zeta \eta} \right) $\\ \hline
    $ \displaystyle {Z}^{-}_{ \mu \nu} {Z}^{-}_{ \rho \sigma} {{(W^+)}}^{+}_{ \xi \tau} {{(W^-)}}^{+}_{ \zeta \eta} \operatorname{Tr} \left( \bar{\sigma}^{\xi \tau} \bar{\sigma}^{\zeta \eta} \right) \operatorname{Tr} \left( \sigma^{\mu \nu} \sigma^{\rho \sigma} \right) $\\ \hline
    $ \displaystyle {Z}^{-}_{ \mu \nu} {Z}^{+}_{ \rho \sigma} {{(W^+)}}^{-}_{ \xi \tau} {{(W^-)}}^{+}_{ \zeta \eta} \operatorname{Tr} \left( \bar{\sigma}^{\rho \sigma} \bar{\sigma}^{\zeta \eta} \right) \operatorname{Tr} \left( \sigma^{\mu \nu} \sigma^{\xi \tau} \right) $\\
\end{longtable}

\subsection{ $W^+ W^+ W^- W^-$ }\label{app:WWWW}\hspace*{\parindent}
$\bullet$ {Dimension = 4, $\mathcal{O}_{4}^{1\sim 2}$ }
\begin{longtable}[c]{|l|l|}
    \hline
    \multicolumn{2}{|c|}{ $W^+ W^+ W^- W^- \quad \pd=4 \quad \mathcal{O}_{4}^{1\sim 2}$} \\ \hline
    \endfirsthead %
    \endhead%
    \hline
    \endfoot%
    $ \displaystyle {{(W^+)}}_{ \mu} {{(W^+)}}_{ \nu} {{(W^-)}}_{ \rho} {{(W^-)}}_{ \sigma} \operatorname{Tr} \left( \sigma^{\mu} \sigma^{\nu} \bar{\sigma}^{\sigma} \bar{\sigma}^{\rho} \right) $ & $ \displaystyle {{(W^+)}}_{ \mu} {{(W^+)}}_{ \nu} {{(W^-)}}_{ \rho} {{(W^-)}}_{ \sigma} \operatorname{Tr} \left( \sigma^{\mu} \bar{\sigma}^{\rho} \right) \operatorname{Tr} \left( \sigma^{\nu} \bar{\sigma}^{\sigma} \right) $\\ 
\end{longtable}
$\bullet$ {Dimension = 6, $\mathcal{O}_{6}^{1\sim 18}$ }
\begin{longtable}[c]{|l|}
    \hline
    \multicolumn{1}{|c|}{ $W^+ W^+ W^- W^- \quad \pd=6 \quad \mathcal{O}_{6}^{1\sim 18}$} \\ \hline
    \endfirsthead %
    \endhead%
    \hline
    \endfoot%
    $ \displaystyle {{(W^+)}}_{ \mu} {{(W^+)}}_{ \nu} {{(W^-)}}^{+}_{ \rho \sigma} {{(W^-)}}^{+}_{ \xi \tau} \operatorname{Tr} \left( \sigma^{\mu} \sigma^{\nu} \bar{\sigma}^{\xi \tau} \bar{\sigma}^{\rho \sigma} \right) $\\ \hline
    $ \displaystyle {{(W^+)}}^{+}_{ \mu \nu} {{(W^+)}}_{ \rho} {{(W^-)}}_{ \sigma} {{(W^-)}}^{+}_{ \xi \tau} \operatorname{Tr} \left( \bar{\sigma}^{\mu \nu} \bar{\sigma}^{\sigma} \sigma^{\rho} \bar{\sigma}^{\xi \tau} \right) $\\ \hline
    $ \displaystyle {{(W^+)}}^{+}_{ \mu \nu} {{(W^+)}}^{+}_{ \rho \sigma} {{(W^-)}}_{ \xi} {{(W^-)}}_{ \tau} \operatorname{Tr} \left( \bar{\sigma}^{\mu \nu} \bar{\sigma}^{\rho \sigma} \bar{\sigma}^{\tau} \bar{\sigma}^{\xi} \right) $\\ \hline
    $ \displaystyle {{(W^+)}}^{-}_{ \mu \nu} {{(W^+)}}_{ \rho} {{(W^-)}}_{ \sigma} {{(W^-)}}^{+}_{ \xi \tau} \operatorname{Tr} \left( \sigma^{\rho} \sigma^{\mu \nu} \sigma^{\sigma} \bar{\sigma}^{\xi \tau} \right) $\\ \hline
    $ \displaystyle {{(W^+)}}_{ \mu} {{(W^+)}}_{ \nu} {{(W^-)}}^{-}_{ \rho \sigma} {{(W^-)}}^{+}_{ \xi \tau} \operatorname{Tr} \left( \sigma^{\mu} \sigma^{\rho \sigma} \sigma^{\nu} \bar{\sigma}^{\xi \tau} \right) $\\ \hline
    $ \displaystyle {{(W^+)}}^{-}_{ \mu \nu} {{(W^+)}}^{+}_{ \rho \sigma} {{(W^-)}}_{ \xi} {{(W^-)}}_{ \tau} \operatorname{Tr} \left( \bar{\sigma}^{\rho \sigma} \bar{\sigma}^{\xi} \sigma^{\mu \nu} \bar{\sigma}^{\tau} \right) $\\ \hline
    $ \displaystyle {{(W^+)}}_{ \mu} {{(W^+)}}^{+}_{ \nu \rho} {{(W^-)}}^{-}_{ \sigma \xi} {{(W^-)}}_{ \tau} \operatorname{Tr} \left( \sigma^{\mu} \sigma^{\sigma \xi} \bar{\sigma}^{\tau} \bar{\sigma}^{\nu \rho} \right) $\\ \hline
    $ \displaystyle {{(W^+)}}^{-}_{ \mu \nu} {{(W^+)}}^{-}_{ \rho \sigma} {{(W^-)}}_{ \xi} {{(W^-)}}_{ \tau} \operatorname{Tr} \left( \sigma^{\xi} \sigma^{\mu \nu} \sigma^{\rho \sigma} \bar{\sigma}^{\tau} \right) $\\
\end{longtable}
$\bullet$ {Dimension = 8, $\mathcal{O}_{8}^{1\sim 42}$ }
\begin{longtable}[c]{|l|}
    \hline
    \multicolumn{1}{|c|}{ $W^+ W^+ W^- W^- \quad \pd=8 \quad \mathcal{O}_{8}^{1\sim 42}$} \\ \hline
    \endfirsthead %
    \endhead%
    \hline
    \endfoot%
    $ \displaystyle {{(W^+)}}^{+}_{ \mu \nu} {{(W^+)}}^{+}_{ \rho \sigma} {{(W^-)}}^{+}_{ \xi \tau} {{(W^-)}}^{+}_{ \zeta \eta} \operatorname{Tr} \left( \bar{\sigma}^{\mu \nu} \bar{\sigma}^{\rho \sigma} \bar{\sigma}^{\zeta \eta} \bar{\sigma}^{\xi \tau} \right) $\\ \hline
    $ \displaystyle {{(W^+)}}^{-}_{ \mu \nu} {{(W^+)}}^{-}_{ \rho \sigma} {{(W^-)}}^{-}_{ \xi \tau} {{(W^-)}}^{-}_{ \zeta \eta} \operatorname{Tr} \left( \sigma^{\mu \nu} \sigma^{\xi \tau} \sigma^{\zeta \eta} \sigma^{\rho \sigma} \right) $\\ \hline
    $ \displaystyle {{(W^+)}}^{+}_{ \mu \nu} {{(W^+)}}^{+}_{ \rho \sigma} {{(W^-)}}^{+}_{ \xi \tau} {{(W^-)}}^{+}_{ \zeta \eta} \operatorname{Tr} \left( \bar{\sigma}^{\mu \nu} \bar{\sigma}^{\xi \tau} \right) \operatorname{Tr} \left( \bar{\sigma}^{\rho \sigma} \bar{\sigma}^{\zeta \eta} \right) $\\ \hline
    $ \displaystyle {{(W^+)}}^{+}_{ \nu \rho} {{(W^+)}}^{+}_{ \sigma \xi} {{(W^-)}}_{ \tau} \left( D_{\mu} {{(W^-)}}^{+}_{ \zeta \eta} \right) \operatorname{Tr} \left( \bar{\sigma}^{\nu \rho} \bar{\sigma}^{\tau} \bar{\sigma}^{\mu} \bar{\sigma}^{\sigma \xi} \bar{\sigma}^{\zeta \eta} \right) $\\ \hline
    $ \displaystyle {{(W^+)}}^{-}_{ \mu \nu} {{(W^+)}}^{-}_{ \rho \sigma} {{(W^-)}}^{+}_{ \xi \tau} {{(W^-)}}^{+}_{ \zeta \eta} \operatorname{Tr} \left( \bar{\sigma}^{\xi \tau} \bar{\sigma}^{\zeta \eta} \right) \operatorname{Tr} \left( \sigma^{\mu \nu} \sigma^{\rho \sigma} \right) $\\ \hline
    $ \displaystyle {{(W^+)}}^{-}_{ \mu \nu} {{(W^+)}}^{+}_{ \rho \sigma} {{(W^-)}}^{-}_{ \xi \tau} {{(W^-)}}^{+}_{ \zeta \eta} \operatorname{Tr} \left( \bar{\sigma}^{\rho \sigma} \bar{\sigma}^{\zeta \eta} \right) \operatorname{Tr} \left( \sigma^{\mu \nu} \sigma^{\xi \tau} \right) $\\ \hline
    $ \displaystyle {{(W^+)}}^{+}_{ \mu \nu} {{(W^+)}}^{+}_{ \rho \sigma} {{(W^-)}}^{-}_{ \xi \tau} {{(W^-)}}^{-}_{ \zeta \eta} \operatorname{Tr} \left( \bar{\sigma}^{\mu \nu} \bar{\sigma}^{\rho \sigma} \right) \operatorname{Tr} \left( \sigma^{\xi \tau} \sigma^{\zeta \eta} \right) $\\
\end{longtable}